\documentclass[aps,prb,twocolumn,superscriptaddress,showpacs]{revtex4-1}

\usepackage{graphicx}
\usepackage{color}
\usepackage{amsbsy,amssymb,amsmath,bm}
\usepackage{float}
\usepackage{verbatim}
\usepackage[backref,colorlinks,linkcolor=blue,citecolor=red,pagecolor=red]{hyperref}

\newcommand{\e}{\varepsilon}
\newcommand{\mubar}{\bar{\mu}}
\newcommand{\vavr}{\langle|\vec{v}|\rangle}

\newcommand{\ehat}{\hat{e}}
\newcommand{\shat}{\hat{s}}
\newcommand{\vhat }{\hat{v}}

\newcommand{\qhat }{\hat{q}}

\begin{document}

\title{Thermodynamics and the Quantum Transport of Particles and Entropy}

\author{C. Strunk}
\affiliation{Institute of Experimental and Applied Physics, University of Regensburg, D-93025 Regensburg, Germany}
\email[]{christoph.strunk@ur.de}

\date{\today}

\begin{abstract}
A unified view on macroscopic thermodynamics and quantum transport is presented. Thermodynamic processes with an exchange of energy between two systems necessarily involve the flow of other {\it balanceable} quantities. These flows are first analyzed using a simple drift-diffusion model, which includes the thermoelectric effects, and connects the various transport coefficients to certain thermo\-dynamic susceptibilities and a diffusion coefficient.
In the second part of the paper the connection between macroscopic thermodynamics and quantum statistics is discussed. It is proposed to employ not particles, but {\it elemen\-tary Fermi- or Bose-systems} as the elementary building blocks of ideal quantum gases. In this way, the transport not only of particles, but also of entropy can be derived in a concise way, and is illustrated both for ballistic quantum wires, and for diffusive conductors.
In particular, the quantum interference of entropy flow is in close correspondence to that of electric current.
\end{abstract}


\maketitle
\flushbottom
\section{Introduction}

Usually, macroscopic thermodynamics is derived on the basis of statistical mechanics and hence considered to be of essentially  mechanical origin. This review provides a complementary point of view. It takes thermodynamics as the starting point and presents an elementary and transparent description of the quantum transport of particles and entropy, both in the ballistic and the diffusive limit. Many results, scattered throughout the literature, are re-derived within a single, coherent approach.

As a heritage of the classical view on physics, particles interacting via certain force fields are considered as the elementary building blocks of matter. The evolution of quantum mechanics has forced us to accept that this statement can only be kept valid, if the meaning of the word 'particle' is considerably modified: rather than being a point-like classical object with well defined position an momentum, particles of any kind have been understood as the quantized excitations of different quantum fields. In this way a unified description of fields and matter has been achieved, which is known as 2$^{nd}$ quantization.
On the other hand, physics education still has to start out in the classical world. This leaves a deep imprint in our mind and language. When we want to proceed into the regime of quantum physics this imprint is not easy to overcome. This becomes particularly obvious in the field of quantum transport.

The transport of physical quantities like energy, momentum, angular momentum, electrical charge or entropy ($E,\vec{P},\vec{L},Q$, or $S$) can be described in terms of the transport of suitably chosen particles or quasiparticles, which carry the other physical quantities in the sense that with each particle a certain amount of $E,\vec{P},\vec{L},Q$, or $S$ is associated, which is transported along with the particles.
 This means that the transport of particles is unavoidably connected to the transport of other physical quantities. Importantly, only those physical quantities can be transported for which a {\it balance} is possible, which tells us which amount of this quantity has left at time $t_1$ a certain volume element $V_1$ in space and has arrived at a time $t_2$ in another volume element $V_2$ in space. The prototype of such balances are those of {\it amounts} of a substance, of goods or of money.  It is suggested to call the physical quantities allowing similar operations {\it balanceable} or {\it substance-like}. 
A local density $x$, and a local current density $\vec{j}_X$ can be associated with each balanceable quantity $X$. This property allow to formulate conservation laws in term of a continuity equation for each balanceable quantity (see below).

Note that balanceable quantities are not necessarily conserved: important examples are entropy, or the spin. The total angular momentum is of course conserved, but the spin of the moving particles under consideration can be transferred to other systems, i.e., by spin-dependent scattering processes. Entropy can be generated, without extracting it from another system. Particles like photons, phonons or excitons can be created or annihilated, provided that the canonic conservation laws for energy, momentum, and angular momentum are obeyed.

Although the description of transport in terms of the motion of particles can be very easily visualized, it is hard to avoid the traps of classical physics in doing so. The reason is that quantum properties, in particular the indistinguishability of identical particles, have no counterpart in the classical world, although they clearly show up in the thermodynamic and transport properties of matter at the macroscopic level. The goal of this article is to formulate a description of transport processes in systems of indistinguishable particles, which does {\it not} contain elements incompatible with the statistical concepts of quantum physics.

The building blocks of this description are the {\it elementary Bose}- or {\it Fermi-Systems} introduced below. Interestingly, the nature of these systems is microscopic and macroscopic at the same time. They are not point-like, but (in the same sense as a wave function) spatially extended,  they obey the laws of macroscopic or phenomenological thermodynamics, and the quantum mechanical average value of their particle number can be {\it small} -- i.e., down to \textit{one} or even less. In conventional language these systems are termed 'single particle states' and can be populated by one (in the case of fermions) or many (in the case of bosons) particles. The advantage of introducing such systems -- which have no place in a classical world -- is the following: they allow to avoid a terminology, according to which non- or weakly interacting quantum particles occupy the states of a 'one-particle system', which is described by the conventional Schr\"{o}dinger equation. Although this terminology is well established, it leads to a confusion of the concepts of 'state' and 'system', which is very disadvantageous if one tries to formulate a consistent thermodynamic description of systems with identical particles.

In the course of the development of thermodynamics it became apparent that the traditional concept of {\it heat} had to be split into two more abstract concepts:\cite{tisza} one, {\it entropy}, which is specific for thermal phenomena, and a second one,  {\it energy}, which is relevant in all branches of physics. The terms 'heat' and 'heat current' survived in modern physics as a concept characteristic for {\it processes}, namely the amount of energy transferred from one system into another together with a given amount of entropy. This conception of heat, however, is problematic: despite the fact that heat cannot be balanced,\cite{note00} nor linked to quantum states,\cite{note0} it often competes with the much more powerful concept of entropy. Even today,\cite{dubi2011}  entropy is less popular than heat, possibly because of persistent tradition, and its incompatibility with classical mechanics.\cite{note_3rd_law} On the other hand, entropy can be handled very easily at the macroscopic level, as it behaves as an analogue of electric charge (except for the property of conservation). In this article, we take the existence of entropy as a starting point for thermodynamics -- in the same way, as one takes the existence of electric charge as  starting point for electricity. Like in electrodynamics, we don't ask what electric charge or entropy actually 'are', but take them as fundamental concepts, which prove useful in the quantitative description of electric or thermal phenomena, respectively. An intuition for both charge and entropy can be developed only via the many examples, where we see them 'at work'. Our approach allows a natural integration of the ideas of quantum physics.

This article is organized as follows: in section~\ref{sec_thermo} we first formulate thermodynamics in a self-contained way that is appropriate for the investigation of transport processes. In section~\ref{sec:drift-diffusion} we present the simplest of all transport theories, the drift-diffusion model, which has the great advantage of providing a simple intuitive picture of diffusive transport. In this approach particles and entropy are treated on the same footing. In section~\ref{sec_statistics}, we derive quantum statistics by combining thermodynamics with elementary ideas of quantum physics. In section~\ref{sec_el_syst} we introduce {\it elementary Fermi-and Bose-systems} as the elementary building blocks of quantum gases, and derive their thermodynamic equations-of-state. In sections~\ref{sec_Q-transport} and~\ref{sec:S-transport} we apply these equations-of-state to ballistic transport in one-dimensional quantum wires in the Landauer-B\"{u}ttiker approach. In section~\ref{sec:boltzmann_eq} we show that the very same equations-of-state can be applied to generalize the drift-diffusion model in a way, which is equivalent to the Boltzmann equation in relaxation time approximation. From this perspective, there is no fundamental difference between classical and quantum transport. In section~\ref{sec:discussion} some implications of our approach are discussed -- in particular it is shown that also quantum interference can be included into the discussion of thermal transport phenomena.

\section{Thermodynamics}\label{sec_thermo}
As pioneered by Massieu and Gibbs, thermodynamics can be based on the postulate that the static properties of any physical system with $r$ independent variables can be compressed into certain functions of these variables, which are called \textit{thermodynamic potentials}.\cite{callen} The most familiar thermodynamic potential is the energy $E$ when expressed as a function of the independent {\it extensive} variables of the system. In the case of simple fluid or gas, the independent extensive variables are the entropy~$S$, the volume~$V$ and the particle number~$N$. Assigning values for a set of independent variables, e.g., for $\{S,V,N\}$, specifies a certain {\it state} of the system. The total differential of the function $E(S,V,N)$ can then be written in the form
\begin{equation}\label{gl_GFF}
dE\ =\ T\,dS - p\,dV+\mu\,dN\;,
\end{equation}
where the absolute temperature~$T$ is defined as
\begin{equation}\label{gl_Z_th}
T(S,V,N)\ =\ \frac{\partial E(S,V,N)}{\partial S}\;,
\end{equation}
 the pressure~$p$ as
\begin{equation}\label{gl_Z_mech}
-p(S,V,N)\ =\ \frac{\partial E(S,V,N)}{\partial V}\;,
\end{equation}
 and the chemical potential~$\mu$ as
\begin{equation}\label{gl_Z_chem}
\mu(S,V,N)\ =\ \frac{\partial E(S,V,N)}{\partial N}\;.
\end{equation}
Equations~\ref{gl_Z_th}, \ref{gl_Z_mech}, and \ref{gl_Z_chem} can be called
the caloric, thermal and chemical {\it equation of state} (EoS) of the system, and they define the to
$S$, $V$, and $N$ thermodynamically conjugate variables $T$, $p$, and $\mu$. The EoS constitute a complete characterization of a specific system, and their knowledge is (up to an integration constant) equivalent to that of $E(S,V,N)$. In mechanics or electrostatics a potential serves the purpose of combining three force or electric field components into a single function (the potential energy). In the very same way the energy (and also other thermodynamic potentials)  combines the information contained in the three EoS into the single function $E(S,V,N)$.

Equation~\ref{gl_GFF} is called the Gibbs fundamental form (GFF) and has a very simple physical meaning: the intensive quantities $T$, $-p$, and $\mu$ tell how much energy has to be added to or removed from the system, if the extensive quantities $S$, $V$, and $N$ are changed. In many cases a change of the extensive variables corresponds to a transport process: if the system is heated by connecting it to an external reservoir for energy and entropy while $V$ and $N$ are kept constant, the amount $\Delta E=T\Delta S$ of energy together with the amount $\Delta S$ of entropy has to be transported from the reservoir into the system. If  particles are added to the system at constant $S$ and $V$ from an external container, in addition to the amount $\Delta N$ of particles the amount $\mu \Delta N$ of energy has to be transferred from the reservoir to the system.

Depending on the specific problem under consideration, it is often convenient to use a different set of independent variables by exchanging any of the extensive variables $\{S, V, N\}$ with its thermodynamically conjugate partner. If, e.g., $\{T,V,N\}$ are chosen as set of independent variables the corresponding thermodynamic potential is the free energy $F(T,V,N)=E(T,V,N)-T\cdot S(T,V,N)$. In the following the grand canonical, or Landau potential
\begin{equation}\label{gl_K_landau}
K(T,V,\mu)=E-TS-\mu N
\end{equation}
 will play a primary role, which is adapted to $\{T,V,\mu\}$ as set of independent variables.

The fact that the EoS can be derived from a thermodynamic potential implies that they are not independent of each other. As the mixed $2^{nd}$-order partial derivatives of continuously differentiable functions are equal, there must exist many relations between the partial derivatives of the different EoS. With $\{T,V,N\}$ as set of independent variables one finds, e.g.:
\begin{multline}\label{gl_maxwell}
\frac{\partial S(T,V,N)}{\partial V}
=-\frac{\partial^2F(T,V,N)}{\partial V\partial
T}\\=\ -\frac{\partial^2F(T,V,N)}{\partial T\partial V}=\frac{\partial
p(T,V,N)}{\partial T}\;.
\end{multline}
Relations of this type are called {\it Maxwell-Relations}.\cite{note_maxwell}

The final feature of thermodynamics needed for the present work is the homogeneity postulate:\cite{footnote3} it is assumed that the thermodynamic potentials $Y(X_1,\dots X_j,\xi_{j+1},\dots,\xi_r)$ of any system have to be {\it homogeneous} in the extensive variables $\{X_1,\dots,X_j\}$, i.e.
\begin{eqnarray}\nonumber
\lambda\,Y(X_1,\dots &X_j,&\xi_{j+1},\dots,\xi_r)
\\\ &=&\ Y(\lambda X_1,\dots,\lambda X_j,\xi_{j+1},\dots,\xi_r)\;,\label{gl_hom1}
\end{eqnarray}
where $\{\xi_{j+1},\dots,\xi_r\}$ are the independent intensive quantities, and $\lambda$ is an arbitrary dimensionless scaling factor. Equation~\ref{gl_hom1} implies the Euler or homogeneity relation
\begin{equation}\label{gl_hom2}
E\ =\ TS -pV +\mu N\;.
\end{equation}
The physical meaning of the homogeneity\cite{footnote4} is quite fundamental: it implies that all physical systems obey a scaling relation, which expresses the relations between their state variables in a way that is independent of the 'size' of the system, i.e., all properties of the system can be expressed by relations between the {\it intensive} quantities $e, s, n, T$, and $\mu$ .

 The homogeneity of $E(S,V,N)$, for example, means that $E$ can be written as
\begin{equation}\label{gl_hom3}
E(S,V,N)\ =\ V\cdot e(s,n)\;,
\end{equation}
where $e=E/V$ is the energy density, $s=S/V$ the entropy density, and $n=N/V$ the particle density.
This implies that the function $e(s,n)$ represents a {\it reduced} thermodynamic potential, which still contains all thermodynamic information about the system, except its volume.  The corresponding reduced Gibbs fundamental form reads
\begin{equation}\label{red_GFF1}
de\ =\ T\,ds\;+\,\mu\,dn\;.
\end{equation}

\noindent Equally well one can rewrite Eq.~\ref{gl_hom3} as
\begin{equation}\label{gl_hom3b}
E(S,V,N)\ =\ N\cdot \ehat(\shat,\vhat )\;,
\end{equation}
where $\ehat=E/N$ is the energy per particle, $\shat=S/N$ the entropy per particle, and $\vhat =1/n=V/N$ the volume per particle.
Then the function $\ehat(\shat,\vhat )$ represents another reduced thermodynamic potential, which contains all thermodynamic information about the system, except its particle number. The corresponding reduced GFF reads
\begin{equation}\label{red_GFF2}
d\ehat\ =\ T\,d\shat\;-\,p\,d\vhat \;,
\end{equation}
which is the form implicitly considered in many textbooks, after stating that $N$ is assumed to be fixed. The latter choice is preferred in physical chemistry, where it allows to express the specific properties of a substance in a given aggregation state in a way that is independent of the amount $N$ of the substance.

If we choose $e(s,n)$ as reduced thermodynamic potential, we can still apply the formalism of Legendre transforms to exchange the independent variables, i.e., $s$ with $T$ and $n$ with $\mu$. If we do so, we find using Eq.~\ref{gl_hom1}
\begin{equation}\label{gl_pressure}
-p(T,\mu)\ =\ e(T,\mu)-T\cdot s(T,\mu)-\mu\cdot n(T,\mu)
\end{equation}
as the corresponding reduced thermodynamic potential. The differential of $p(T,\mu)$ represents a {\it reduced} fundamental form, which is also known as the Gibbs-Duhem relation
\begin{equation}\label{gl:gibbs_duhem}
dp\ =\ s\,dT +n\,d\mu\;,
\end{equation}
with the two equations of state
\begin{equation}\label{gl_Z_loc}
s(T,\mu)\;=\;\frac{\partial p(T,\mu)}{\partial T}\quad\text{and}\quad
n(T,\mu)\;=\;\frac{\partial p(T,\mu)}{\partial \mu}\;.
\end{equation}
Using Eq.~\ref{gl_hom2}, we see that the reduced thermodynamic potential $-p(T,\mu)$ is equivalent to the Landau potential
\begin{equation}\label{gl_hom4}
K(T,V,\mu)\ =\ -V\,p(T,\mu)\;.
\end{equation}

\begin{figure}[t]
\centering
\includegraphics[width=70mm]{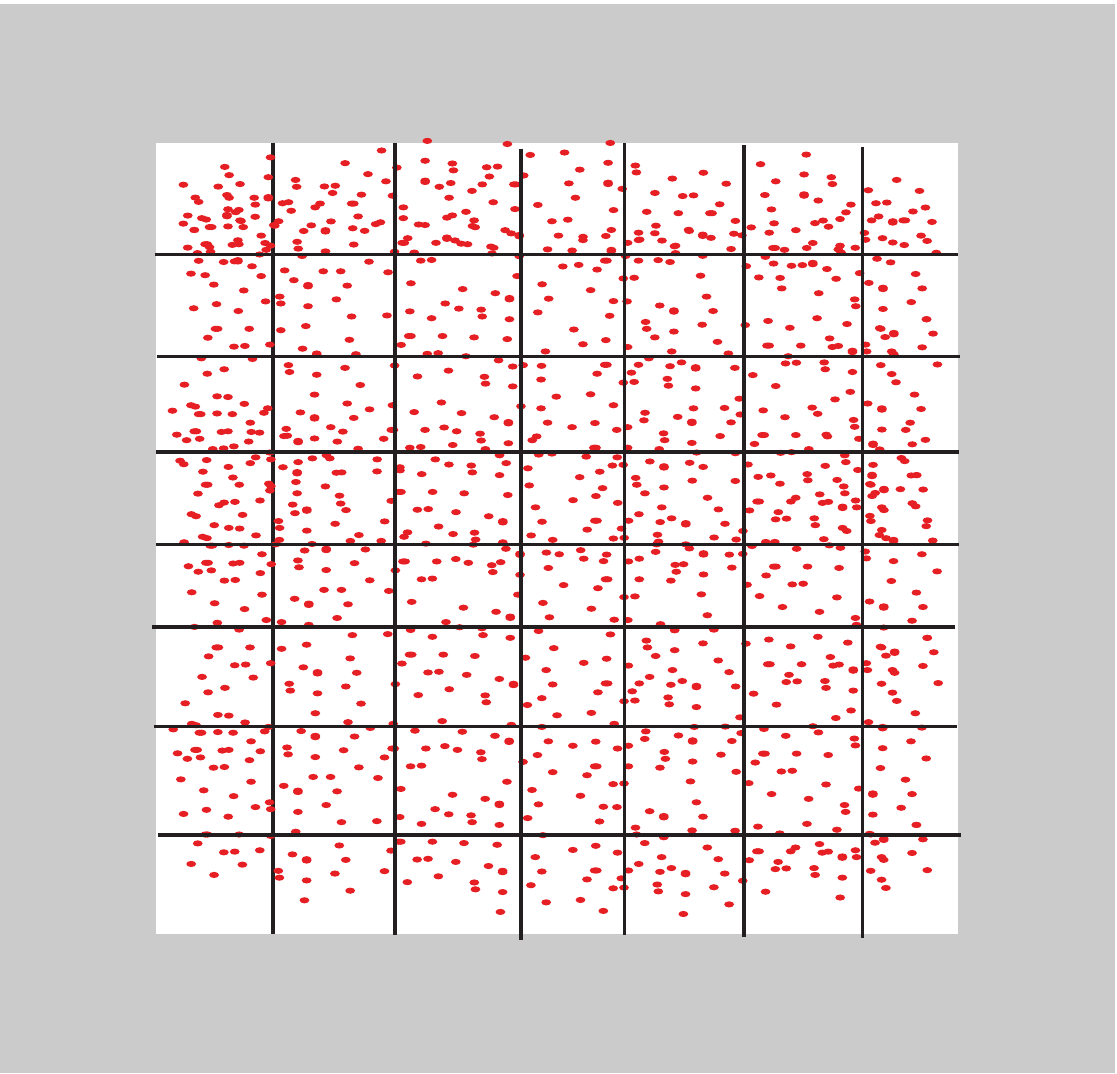}
\caption[]{A container with a gas of (quasi)-particles can be decomposed into small subvolumina with arbitrary size. Each subvolume represents another realization of the system 'gas', which continuously exchanges energy, entropy and particles with its neighbors. In presence of a gradient of $T$ or $\mu$ each subvolume can be still considered to be in local  equilibrium, provided that its volume is not smaller than $\Lambda^3$ ($\Lambda$ is the mean free path between scattering events, see Sec.~\ref{sec:drift-diffusion}). The state of each subvolume is then characterized by the {\it local} values of $T$ and $\mu$. }\label{fig:particles_in_bath}
\end{figure}

The choice of $\{T,V,\mu \}$ as independent variables perfectly matches the needs for a description of an important class of non-equilibrium situations.
If we decompose a macroscopic solid, liquid or gas into small volume elements as it is illustrated in Fig.~\ref{fig:particles_in_bath}, assuming that each of these volume elements is in local thermodynamic equilibrium, we can try to model situations where the local temperature and/or the local chemical potential are spatially varying. In this case the gradients of $T$ and $\mu$ are known to lead to the diffusion of energy, entropy and particles.

It is appealing that also external fields like the gravitational field or the electrostatic field can be built into the local thermodynamics. In a system with electrically charged particles we have one more extensive variable, i.e., the electric charge $Q$, which provides an extra term in the Gibbs fundamental form (see Eq.~\ref{gl_GFF}):
\begin{equation}\label{gl_GFFQ}
dE\ =\ T\,dS - p\,dV+\mu\,dN+\phi\,dQ\;.
\end{equation}
The to $Q$ thermodynamically conjugate variable is the electrostatic potential $\phi$, which determines the electrostatic contribution to the energy required for a local increase of the charge density.

Usually charge and particle number are connected by a characteristic constant of the system, i.e., the charge per particle $\hat{q}$.\cite{footnote5} In these cases charge and particle number are not independent, but proportional: $Q=\hat{q}N$, implying that we can combine the last two terms in Eq.~\ref{gl_GFFQ}
\begin{equation}\label{gl_GFFQ2}
dE\ =\ T\,dS - p\,dV+\mubar\,dN\;
\end{equation}
where
\begin{equation}\label{gl_mubar}
\mubar\ :=\ \mu+\hat{q}\phi
\end{equation}
defines the {\it electrochemical potential}. For charged particles it is $\mubar$ and not $\mu$, which quantifies the energy changes required for adding or removing particles. Hence $\mubar$ and not just $\mu$ enters all thermodynamic relations for systems of charged particles.\cite{footnote6}

\begin{figure}[t]
\centering
\includegraphics[width=30mm]{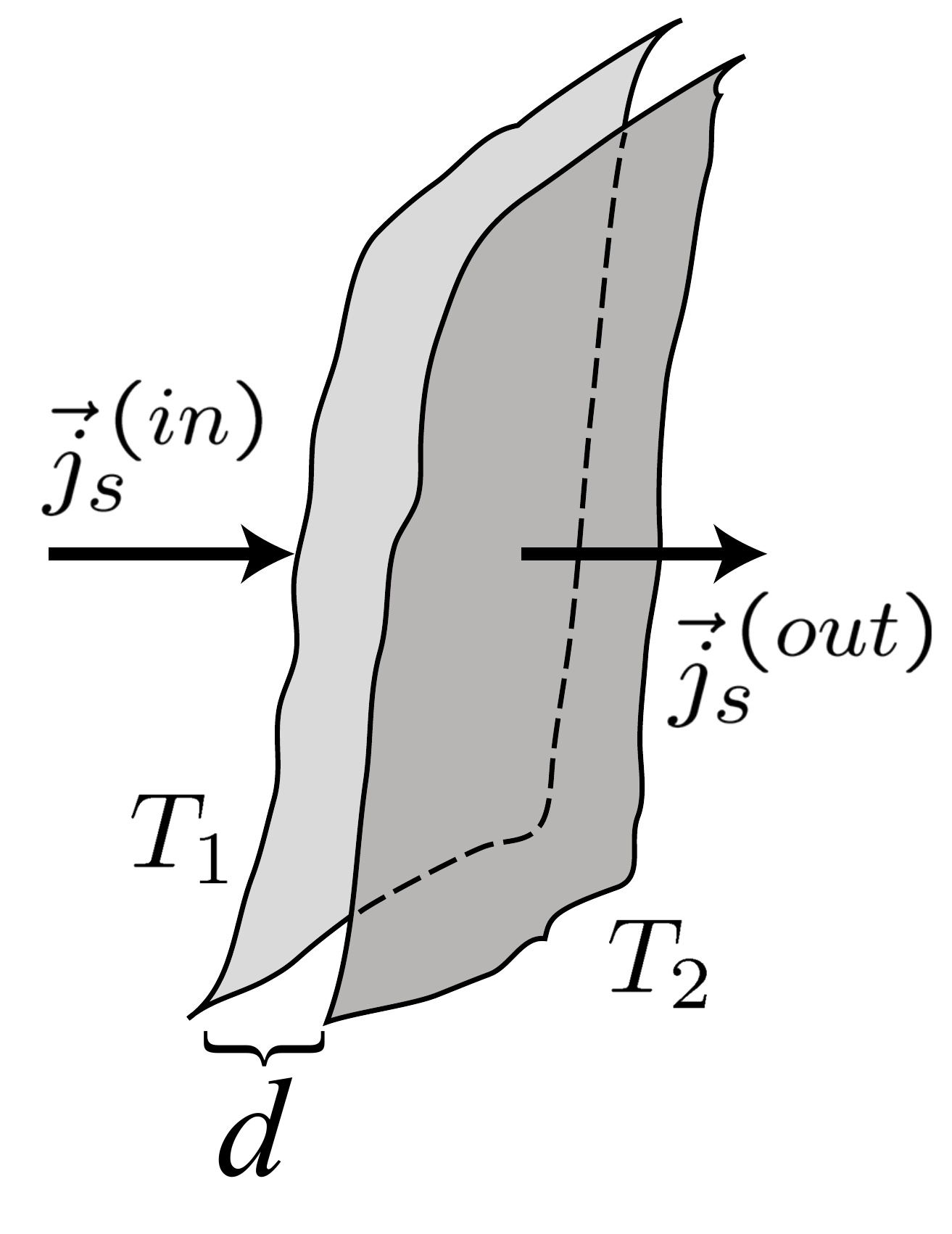}
\caption{An entropy current flowing through two isothermal surface elements with temperatures $T_1$ and $T_2\lesssim T_1$.
The total rate of entropy production between the two surfaces becomes negligible compared to the entropy currents through the surface elements, as the distance $d$ and the temperature difference $T_1-T_2$ go to zero.}
\label{fig:skizze_GGF}
\end{figure}
So far we have not exploited any of the conservation laws.\cite{footnote7} In general, a balanceable quantity $X$ has to obey a {\it continuity equation}
\begin{equation}\label{continuity}
\frac{\partial x(t,\vec{r})}{\partial t}+\nabla\cdot\vec{j}_X(t,\vec{r})\ =\ \Sigma_X(t,\vec{r})\;,
\end{equation}
where $x$ is the local $X$-density, $\vec{j}_X$ the $X$-current density and $\Sigma_X$ the $X$-generation rate per volume. For conserved quantities $\Sigma_X$ vanishes.
Applying the continuity equation for $E$, $S$, and $N$ we can rewrite the reduced GFF (Eq.~\ref{red_GFF1}) in a very intuitive form: the rate of energy transfer $\dot{e}\big(s(t,\vec{r}),n(t,\vec{r})\big)$ from one side of an infinitesimally small surface element in space to the other side reads
\begin{equation}\label{gl_GFF_rate}
\frac{\partial \,e\big(s(t,\vec{r}),n(t,\vec{r})\big)}{\partial t}\ =\ T\,\frac{\partial s}{\partial t}+\mubar\;\frac{\partial n}{\partial t}\;.
\end{equation}
For a stationary flow, where all current densities are constant in time, the time derivatives of $e$, $s$, and $n$ can be replaced by the corresponding current densities
\begin{equation}\label{gl_GFF_J}
\vec{j}_E\ =\ T\,\vec{j}_S\ +\ \mubar\,\vec{j}_N\;,
\end{equation}
because the total generation (or annihilation) rate of entropy and particles on the surface separating two volume elements vanishes, as its volume is zero. This is illustrated in Fig.~\ref{fig:skizze_GGF}.

Equation~\ref{gl_GFF_J} provides a general relation between the energy current density and the current densities of the other independent balanceable quantities of the system. It is as fundamental as Eq.~\ref{gl_GFF}, and should hold for any system in local thermodynamic equilibrium, for which the average velocity $\langle \vec{v}\rangle$ is negligible.
If local equilibrium is maintained in systems with non-negligible average velocity, the (thermodynamically) conjugate variable pair velocity $\langle\vec{v}\rangle$ and momentum $\langle\vec{P}\rangle$ gives rise to another term in Eqs.~\ref{gl_GFF} and \ref{gl_GFF_J} -- including this term would lead to hydrodynamics, and is beyond the scope of this article.
For the diffusive transport to be studied in the next section  $\langle\vec{v}\rangle$ is usually so small that this additional term can be neglected.

On a surface of constant $T$ and $\mubar$ an analogous expression holds for the currents through this surface:
\[I_E\ =\ T\;I_S+\mubar\,I_N\;.\]
This expression tells the strength of the energy current that is 'carried' by the currents $I_S$ and $I_N$, namely $ T\;I_S$, and $\mubar\,I_N$, respectively.\cite{falk}
The term $T\,I_S$ in this relation is usually called the 'heat current', but we have to stress again that 'heat' is not a state variable, to which one can assign a value in an equilibrium state. In particular, one cannot talk of a 'heat content' of the system, despite the fact that its existence is strongly suggested by the terms 'heat current' and 'heat capacity'. For this reason, it is preferable to talk about the entropy content of a system, its current and its local density, where no such conceptual problems exist.\cite{dubi2011}

\section{Drift-Diffusion model}\label{sec:drift-diffusion}
From the point of view of macroscopic thermodynamics, it is natural to describe non-equilibrium situations by spatially varying thermodynamic variables, i.e., the local values of the densities and the intensive variables. An obvious question is, what is the length scale, below which the concepts of a local temperature and a local chemical potential are no longer applicable? It is natural to identify this length scale with the mean free path $\Lambda$ of the scattering processes, which are responsible for establishing local equilibrium. For smaller distances it is not possible to assign a $T$-  or $\mu$-difference. In the simplest case, the mean free path is given by
\begin{equation}\label{gl_lambda}
\Lambda\ =\ \frac{1}{n_{\text{scatt}}\,\sigma_c}\;,
\end{equation}
where $n_\text{scatt}$ is the density of scatterers, and $\sigma_c$ is the cross section of the relevant scattering process.\cite{footnoteZ}

\begin{figure}[t]
\includegraphics[width=66mm]{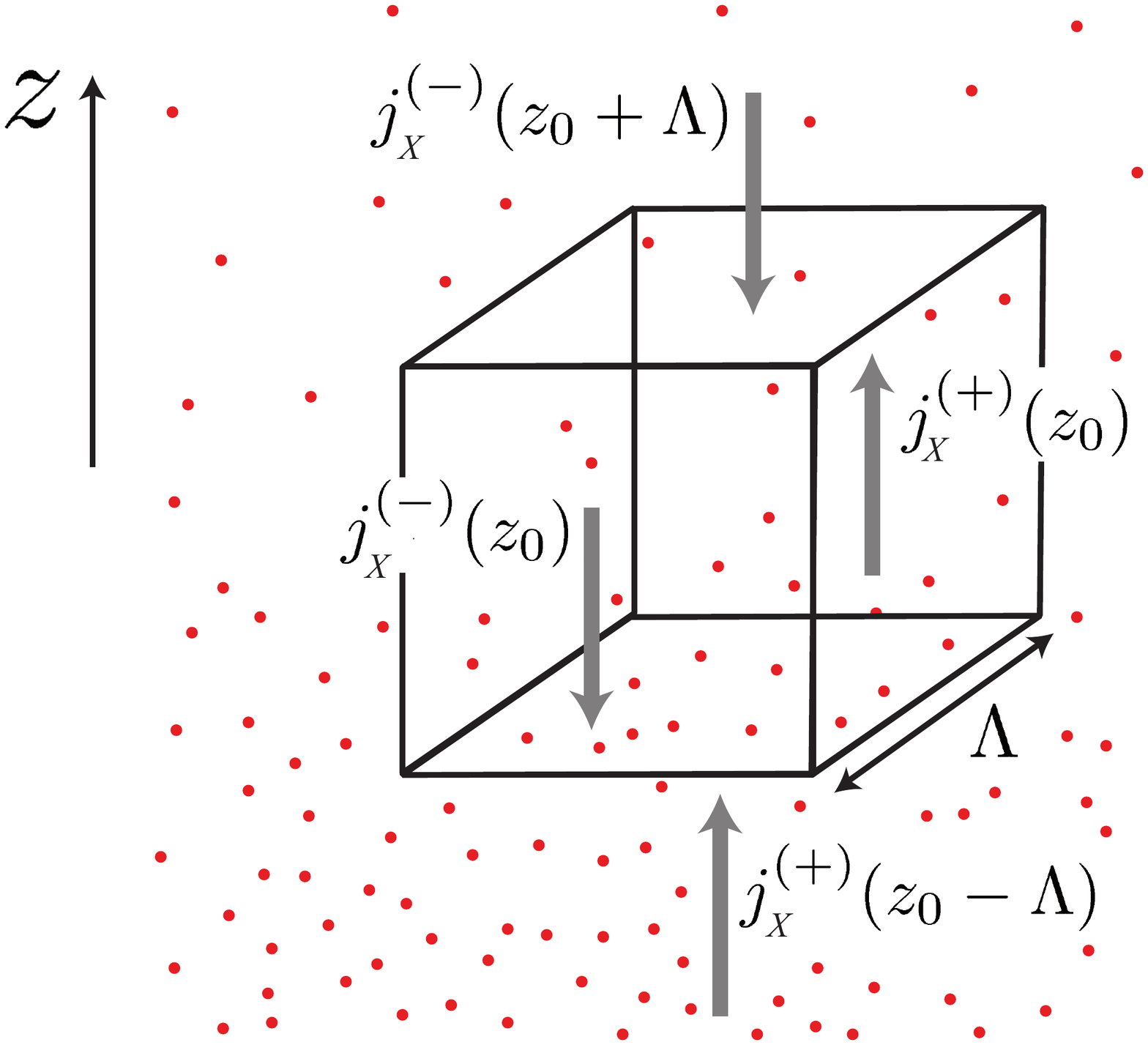}
\caption{\label{fig:diffusion}
Elementary derivation of the diffusion constant: summing up the four contributions
$j_{Xz} =\pm \frac{1}{6}\,x(z)\cdot\langle|\vec{v}|\rangle$
 to the $z$-component $j_{Xz}$ of the $X$-current density  through the top and bottom surface of a cube of dimension $\Lambda$, one arrives in linear approximation at Eq.~\ref{gl_DDmodel}, if the $z$-direction is chosen parallel to the gradient $\nabla x$ of the $X$-density.
}
\end{figure}

If the mobile particles in the system move with average velocity $\vavr$ between the scattering events,\cite{note_velocity}
$\Lambda$ can be translated into a scattering time $\tau$ via the relation $\Lambda= \vavr\cdot \tau$.
The elementary consideration illustrated in Fig.~\ref{fig:diffusion} shows that the current density associated with a balanceable quantity $X$ in linear approximation can be written as
\begin{equation}\label{gl_DDmodel}
\vec{j}_X\ =\ -D\cdot \nabla x(t, \vec{r})\;,
\end{equation}
where
\begin{equation}\label{gl:diff_const}
D=\frac{1}{3}\vavr\Lambda
\end{equation}
 is the diffusion constant.\cite{blundell}

 At face value this derivation appears to be based very much on classical physics. Looking more closely, however, the only thing really exploited it the condition that the quantity $X$ is {\it balanceable}, i.e., that it is possible to say what is the net amount of $X$ transported in the direction of the gradient $\nabla x(\vec{r})$ of the $X$-density. For this reason the model is extremely robust and holds in both classical and quantum physics.
Besides on thermodynamics, it relies on only two additional concepts, i.e., the transport velocity $\vavr$, and the mean free path $\Lambda$ between scattering events.
In quantum physics,
\[\vavr(\vec{k})=\frac{1}{\hbar}\;\nabla_{\vec{k}}\e(k)\]
 is the group velocity resulting from the dispersion relation $\e(k)$ of the matter waves, while $\Lambda$ is derived from the quantum mechanical scattering cross section $\sigma_c$ (see Eq.~\ref{gl_lambda}).

In order to deal with particle currents, we set $X=N$, and  obtain
 \begin{equation}\label{gl_DDmodel_N}
\vec{j}_N\ =\ -D\cdot \nabla n(t, \vec{r})\;,
\end{equation}
which is known as Fick's 1.~law.

Let us first consider the entropy transport. If we choose $T$ and $n$ as independent variables,  we can apply Eq.~\ref{gl_DDmodel} to entropy, and obtain the entropy current density, and the thermal contribution to the energy current density:
\begin{eqnarray}\label{gl_j_S}
\nonumber T\cdot\vec{j}_S\ &=&\ -T\cdot D\nabla s(T,n)\\[2mm]
\nonumber&=&\ -T\cdot D\left\{\frac{\partial s(T,n)}{\partial T}\nabla T+\frac{\partial s(T,n)}{\partial n}\nabla n\right\}\\[1mm]
&=&\ -Dc_v\nabla T+ T\cdot \frac{\partial s(T,n)}{\partial n}\ \vec{j}_N\;
\end{eqnarray}
where $c_v=T\partial s(T,n)/\partial T$ is the heat capacity per volume at constant density $n=N/V$.
The first term in Eq.~\ref{gl_j_S} represents Fourier's law, i.e., the {\it conductive} thermal contribution to the energy current with the thermal conductivity
\begin{equation}\label{gl:lambda}
\lambda=Dc_v=\frac{1}{3}n\hat{c}_v\vavr^2\tau\;.
 \end{equation}
 The second term represents the Peltier effect, where at constant $T$ the moving particles result in a {\it convective} thermal contribution to the energy current. The Peltier coefficient connecting this contribution and the electric current $\vec{j}_Q=\hat{q}\vec{j}_N$ then reads\cite{note_j_S_vs j_N}
\begin{equation}\label{gl_Peltier}
\Pi\ =\ \frac{T}{\hat{q}}\,\frac{\partial s(T,n)}{\partial n}\;.
\end{equation}

Applying the transport equation to the electric charge, we have to choose $T$ and $\mubar$ as independent variables. In addition, we have to take into account that even for $\nabla n(t,\vec{r})=0$, a so-called drift current can be present, which is driven by the electric field. Adding this term to the diffusion current (Eq.~\ref{gl_DDmodel_N}), we obtain for the electric current
 \begin{eqnarray}\label{gl_j_Q}
\vec{j}_Q\ &=&\ -\hat{q}D\cdot \nabla n(t, \vec{r})-\sigma\nabla \phi(t, \vec{r})\\[2mm]
\nonumber&=&\ -\hat{q}D\frac{\partial n(T,\mu)}{\partial \mu}\,\nabla\mu-\sigma\,\nabla\phi-\hat{q}D\frac{\partial n(T,\mu)}{\partial T}\,\nabla T \;,
\end{eqnarray}
where $\sigma$ is the electric conductivity.
In global equilibrium $T$ and $\mubar=\mu+\hat{q}\phi$ are constant. In this case
the current has to vanish and one obtains the {\it Einstein relation} between $D$ and $\sigma$:
\begin{equation}\label{gl_Einstein}
\sigma\ =\ \hat{q}^2\frac{\partial n(T,\mu)}{\partial \mu}\cdot D\;.
\end{equation}
Note that in equilibrium $\mu$ and $\phi$ do not need to vanish separately, a phenomenon, which occurs, e.g., in the vicinity of a $pn$-junction in inhomogeneous semiconductors.

The thermodynamic susceptibility
\begin{equation}\label{gl:particle_capacity}
\nu\ =\ \frac{\partial n(T, \mu)}{\partial \mu}\ =\ n^2\kappa_T
 \end{equation}
 is closely related to the isothermal compressibility $\kappa_T=-(1/\vhat)\cdot\partial\vhat (T,p)/\partial p$. I propose to call $\nu$ the {\it particle capacity}, as it tells us how many particles can be added to the system, if the (electro)-chemical potential is raised by a certain amount.\cite{footnoteY}
In systems of charged particles particle capacity is related to the electric capacitance. In low-dimensional conductors $\partial n(T, \mu)/\partial \mu$ can be very small and then provides a contribution to the total inverse capacitance, which cannot be neglected against that of the  geometrical (electrostatic) capacitance. The sum of both contributions determines the ratio between the electric charge on a capacitor and the electro-chemical potential difference between its electrodes. In the context of low-dimensional conductors $\nu$ is also called 'quantum capacitance'.\cite{footnoteX}

 With the Einstein relation we can rewrite Eq.~\ref{gl_j_Q} in the more compact form
 \begin{equation}\label{gl_j_Q2a}
\vec{j}_Q\ =\ -\frac{\sigma}{\hat{q}}\cdot \nabla \mubar(t, \vec{r})-\hat{q}D\frac{\partial n(T,\mu)}{\partial T}\,\nabla T \;.
\end{equation}
We see that a particle current can not only be driven by a gradient of the electro-chemical potential, but also by a temperature gradient. The second term in Eq.~\ref{gl_j_Q2a} describes the Seebeck effect.
To identify the prefactor in front of $\nabla T$ we employ the Maxwell relation resulting from Eqs.~\ref{gl_Z_loc}:
\begin{equation}\label{gl:maxwell}
\frac{\partial n(T,\mu)}{\partial T}\ =\ \frac{\partial s(T,\mu)}{\partial \mu}
\ =\ \frac{\partial s(T,n)}{\partial n}\cdot\frac{\partial n(T,\mu)}{\partial \mu}\;.
\end{equation} Thus we arrive at
 \begin{equation}\label{gl_j_Q2}\nonumber
\vec{j}_Q\ =\ -\frac{\sigma}{\hat{q}}\cdot \nabla \mubar(t, \vec{r})-\sigma{\cal S}\,\nabla T(t, \vec{r}) \;,
\end{equation}
where
\begin{equation}\label{gl_seebeck}
{\cal S}\ =\ \frac{1}{\hat{q}}\,\frac{\partial s(T,n)}{\partial n}\ =\ -\frac{1}{\hat{q}}\,\frac{\partial \mu(T,n)}{\partial T}\;,
\end{equation}
is the Seebeck coefficient, or the thermopower. The second equality in Eq.~\ref{gl_seebeck} corresponds to another Maxwell equation derived from the free energy density $f(T,n)=e(T,n)-Ts(T,n)$. From  Eq.~\ref{gl:maxwell}, and the assumption that the same diffusion coefficient $D$ in Eq.~\ref{gl_DDmodel} holds for both the entropy and the particle current density, it follows that ${\cal S}$ and $\Pi$ obey the Kelvin-Onsager relation\cite{onsager}
\begin{equation}\label{gl_seebeck_2}
\Pi \ =\ T\cdot\,{\cal S}\;.
\end{equation}
Applying the general transport equation Eq.~\ref{gl_DDmodel} to the energy density $e(T,\mu)$ one can easily verify that the model is consistent with Eq.~\ref{gl_GFF_J}, if one uses the homogeneity relation Eq.~\ref{gl_hom2} and Eq.~\ref{gl_Z_loc}.

The main advantage of the drift diffusion model is its extreme simplicity and generality. It does not depend on the nature of the diffusing particles, and works equally well for classical particles, fermions, or bosons. It is also independent of the dispersion relation of the particles under consideration, e.g., electrons, phonons or photons. This means, it can be used for almost all phenomena related to the diffusive transport of quasiparticles occurring in condensed matter physics. Introducing a single phenomenological parameter, the diffusion constant $D$, it relates all transport coefficients to thermodynamic susceptibilities.

That single parameter $D$ is also its main deficiency, because $D$ usually depends on the energy of the diffusing particles. As we will see in Sec.~\ref{sec:boltzmann_eq}, this deficiency can be quite easily cured, if particles with different kinetic energy $\e$ are associated with different subsystems of the gas, having an energy-dependent diffusion constant $D(\e)$. This additional  dependence on energy produces in many cases only a prefactor of order unity.  Hence, the drift-diffusion model works quite well for the qualitative consideration of electric, thermal, and thermoelectric transport phenomena.

The drift diffusion model is easily extended to two species of particles, e.g., electrons and holes in semiconductors. Another topical example is the transport of spin-up  and spin-down electrons in the context of spintronics,\cite{jaro1} and spin-caloritronics.\cite{bauer,jaro2}

\section{Connection to quantum statistics}\label{sec_statistics}
The explanation of the thermodynamic and transport properties of matter is the central goal of statistical physics. In particular, the thermodynamic derivatives entering the transport coefficients in the previous section, can be calculated using the methods of statistical thermodynamics. In the following it is shown that the Gibbs fundamental form (see Eq.~\ref{gl_GFF}), when combined with elementary statistical considerations, reproduces in a simple way the main results of quantum statistical physics.

Our starting point is the grand canonical ensemble, since this allows an exact implementation of the indistinguishability of quantum particles. This approach allows us to use $\{T,V,\mubar\}$ as set of independent variables, which in the previous section turned out as most appropriate for the study of transport phenomena.

Let us assume that the operators ${\cal H}$ and ${\cal N}$  represent energy  and particle number of a given quantum system, and have common eigenstates $|i\rangle$ with the eigenvalues $E_i$ and $N_i$. If the state of the system is a statistical mixture, in which each eigenstate $|i\rangle$ occurs with the probability $W_i$, the average values of  ${\cal H}$ und ${\cal N}$ are given by\cite{note_milena}
 \begin{equation}\label{gl:constraints}
  E\ =\ \langle{\cal H}\rangle =
\sum\limits_i\; E_i\, W_i\;,\quad N\ =\ \langle{\cal N}\rangle = \sum\limits_i\; N_i\, W_i\;,\end{equation}
while the entropy is given by the famous expression
\begin{equation}\label{gl_S_boltzmann}
S\ =\ -k_B \sum\limits_i W_i\, \ln W_i\,.
\end{equation}
Since the probabilities $W_i$ vary continuously, also the averages $E$ and $N$ form a continuum of real numbers, and obey the usual laws of calculus, even if $\langle{\cal N}\rangle\ll1$. Thermodynamics is now considered as a theory of quantum mechanical average values, and the differentials of $E$, $N$ and $S$ read:
 \begin{eqnarray*}
 dE\ &=&\ \sum\limits_i\; E_i\, dW_i, \quad
dN\ = \ \sum\limits_i\, N_i \,dW_i\\
\text{and~~}\; dS\ &=&\ -k_B \sum\limits_i
\;\big(\ln W_i\big) \,dW_i\;.
\end{eqnarray*}
In the last step we have used the normalization $\sum_iW_i=1$ of probabilities, which implies $\sum_idW_i=0$.
Next, we have to determine the probabilities $\{W_i\}$ such that the basis of macroscopic thermodynamics, i.e., the Gibbs fundamental form (Eq.~\ref{gl_GFF}) is obeyed. Assuming a constant volume $V$, we write Eq.~\ref{gl_GFF} in the form
\begin{eqnarray}\label{GFF_stat}
dK(\{W_i\})\ &=&\ dE-T\,dS-\mubar\,dN\ \\
\nonumber &=&\sum\limits_i\,\Big\{E_i+k_BT\ln W_i-\mubar N_i\Big\}\;dW_i\overset{!}{=}0\;.
\end{eqnarray}
This corresponds to an extremalization of the thermodynamic potential $K(\{W_i\};T,V,\mubar)$ with respect to the $W_i$. Introducing the normalization condition for the $\{W_i\}$ via the Lagrange multiplier $\lambda$, we write
\[\sum\limits_i\,\Big\{E_i +  k_BT \ln W_i - \mubar N_i + \lambda \Big\} \,d W_i \overset{!}{=} 0\;.\]
The $W_i$ can now be treated as independent variables, and we obtain
\begin{equation}\label{gl_ln_Wi} \ln W_i = -
\frac{E_i-\mubar
N_i}{k_BT}-\frac{\lambda}{k_BT}\;.\end{equation}
Hence, we have shown that the probabilities have to follow the Gibbs distribution
\begin{equation}\label{gl:Gibbs distribution}
W_i(T,\mubar) =\frac{1}{{\cal Z}(T,\mubar)} \cdot
  \exp\left(-\frac{E_i - \mubar N_i}
{k_BT}\right)\;,\end{equation}
in order to satisfy the Gibbs fundamental form (Eq.~\ref{gl_GFF}).
 The quantity
\begin{equation}\label{grosskan_zustandsumme}{\;\cal Z}(T,V,\mubar) =
\exp\left(\frac{\lambda}{k_BT}\right) = \sum\limits_i\; \exp
\left(-\frac{E_i -\mubar N_i} {k_BT} \right)\;\end{equation}
is the grand canonical partition function. It ensures the correct normalization of the probabilities $W_i$.\cite{note_Z}
${\cal Z}$ depends on volume, as the eigenvalues $E_i(V)$ of ${\cal H}$ typically vary with $V$ - at least via the the allowed values $k_i$ of the components of $k$: $k_i\propto V^{2/3}$.

Next we insert the probabilities into Eq.~\ref{gl_S_boltzmann} and obtain
\begin{eqnarray*}S &=& - k_B \sum\limits_i \,W_i \cdot \underbrace{\left(-
\frac{E_i-\mubar N_i}{ k_BT}-\ln{\cal Z}(T,V,\mubar)\right)}_{\displaystyle=\ln W_i}\\
&=&k_B \left( \frac{E - \mubar N} { k_BT} + \ln{\cal Z}(T,V, \mubar)\right)\,.
\end{eqnarray*}
Using the homogeneity relation Eq.~\ref{gl_hom2} we finally arrive at the thermodynamic potential
\begin{eqnarray}\label{gl_K_vs_ZG}
K(T,V,\mubar)\;&=&\;-k_BT \ln {\cal Z}(T,V, \mubar)\\\nonumber
&=&\; -\;p(T,\mubar)\cdot V\;.\end{eqnarray}
Hence the statistical approach allows us to express the thermodynamics of a quantum system in terms of its grand canonical partition function. The specific properties of the system under consideration enter via the eigenvalues of its quantum observables ${\cal H}$ and ${\cal N}$.

\section{Elementary Fermi- and Bose-Systems}\label{sec_el_syst}
One essential issue of statistical physics is the decomposition of many-body systems into simpler thermodynamic subsystems, allowing the calculation of $K(T,V,\mubar)$.
If we restrict ourselves to the simplest case of gases with non-interacting quantum particles in a rectangular potential box of volume $V$ the corresponding Hamiltonian reads
 \begin{equation}\label{gl_hamiltonian}
 {\cal H}=\sum_k\,\e(k)\,a^\dagger_k a_k\;,
 \end{equation}
 where the index $k$ represents a wave vector, and $\hbar k$ is the momentum of a particle. The function $\e(k)$ is the dispersion relation of these particles and $a_k^\dagger, a_k$ their creation and annihilation operators, respectively. The number operator is simply ${\cal N}_k=a^\dagger_k a_k$.
 When spin is of interest,  its quantum number $\sigma$ can be simply added to the index $k$; otherwise it gives rise to an additional factor of 2 in front of the sum.

This Ansatz is very general, because many interacting systems can be transformed at least approximately into into a Hamiltonian given by Eq.~\ref{gl_hamiltonian}. In many cases residual interactions between the particles can be taken into account as a renormalization of the function $\e(k)$, and a finite life time of the quasiparticle states resulting from scattering processes.
Hence, the Hamiltonian in Eq.~\ref{gl_hamiltonian} is relevant for a very broad class of systems, i.e., all systems with wave-like excitations. Not only conventional gases, but also complex many-body systems such as the lattice excitations of solids,  Fermi- and Luttinger liquids, superfluids and superconductors, or magnons. Most of the quasiparticles of condensed matter physics are at least approximately described by the Hamiltonian in Eq.~\ref{gl_hamiltonian}.

The structure of  Eq.~\ref{gl_hamiltonian} suggest to decompose an ideal gas of particles or quasi-particles into simpler subsystems. Each subsystem is represented by a single term of the sum in Eq.~\ref{gl_hamiltonian}, i.e., $${\cal H}_k=\e(k)\,a^\dagger_k a_k\;.$$ I propose to call these subsystems {\it elementary Fermi- or Bose-systems}, as they cannot be further decomposed into simpler subsystems.
In a rectangular potential well\cite{footnote_a} the elementary Fermi- or Bose-systems share the same volume $V$ and have the following properties:

\begin{enumerate}
\item[\bf{a)}] {\bf Elementary Fermi-systems} have only two eigenstates of $\cal H$ and $\cal N$ with the eigenvalues $ E_i = \{0,  \varepsilon\}$,
and, respectively,  $\{N_i = 0, 1\} $.

The partition function of this system reads according to Eq.~\ref{grosskan_zustandsumme}
\begin{equation}\label{gl_Z_fermi}
{\cal Z}_F(T,V,\mubar) = 1 + \exp \left(- \frac{\varepsilon(k) - \mubar}
{k_BT}\right)\;.\end{equation}


\item[\bf{b)}] {\bf Elementary Bose-systems} have an infinite, but countable number of eigenstates of $\cal H$ and $\cal N$ with the eigenvalues
     $E_i = \{0,  \varepsilon, 2 \varepsilon, 3
 \varepsilon, ...\}$, and $N_i = \{0,  1, 2 , 3 ,...\}$ .

The partition function is a geometric series in this case and reads:
\begin{equation}\label{gl_Z_Bose}
\;{\cal Z}_B(T,V,\mubar) = \frac{1} {1 - \exp \left(- \frac{\varepsilon(k) - \mubar}
{k_BT}\right)}\;.\end{equation}
\end{enumerate}

\noindent From the partition functions we can obtain for the average particle numbers by differentiating ${\cal Z}_{k}(T,V,\mubar)$ with respect to $\mubar$:
\begin{eqnarray}\label{gl_bosefkt}
 N_k(T,V,\mubar)\ &=&
 \ \frac{1 } {\exp \left(\frac{\varepsilon(k) - \mubar} {k_BT}\right) \pm1}
\;.\end{eqnarray}
The $N_k$ are the well-known Fermi- (upper sign) and the Bose-function (lower sign), respectively.
In Fermi systems $N_k$ varies continuously between 0 and 1, while in Bose-systems $N_k$ varies continuously between 0 and $\infty$. For Bose-systems $\mubar$ has to be always smaller than $\e(k)$ - otherwise the particle number of the system diverges. This divergence of $N_k$ at $\e_k=\mubar$ is the origin of Bose-Einstein condensation. It is important to note that the $N_k$ are average values, around which the particle number of the system labelled $\{k\}$ statistically fluctuates, as opposed to the occupation probability of a single-particle state $|k\rangle$ used in Boltzmann theory.\cite{note_fluct} The average particle numbers $N_k$ are often called {\it distribution functions}, since they tell how the total number of particles is distributed over the different elementary sub-systems.

\begin{figure}[t]
\centering
\includegraphics[width=85mm]{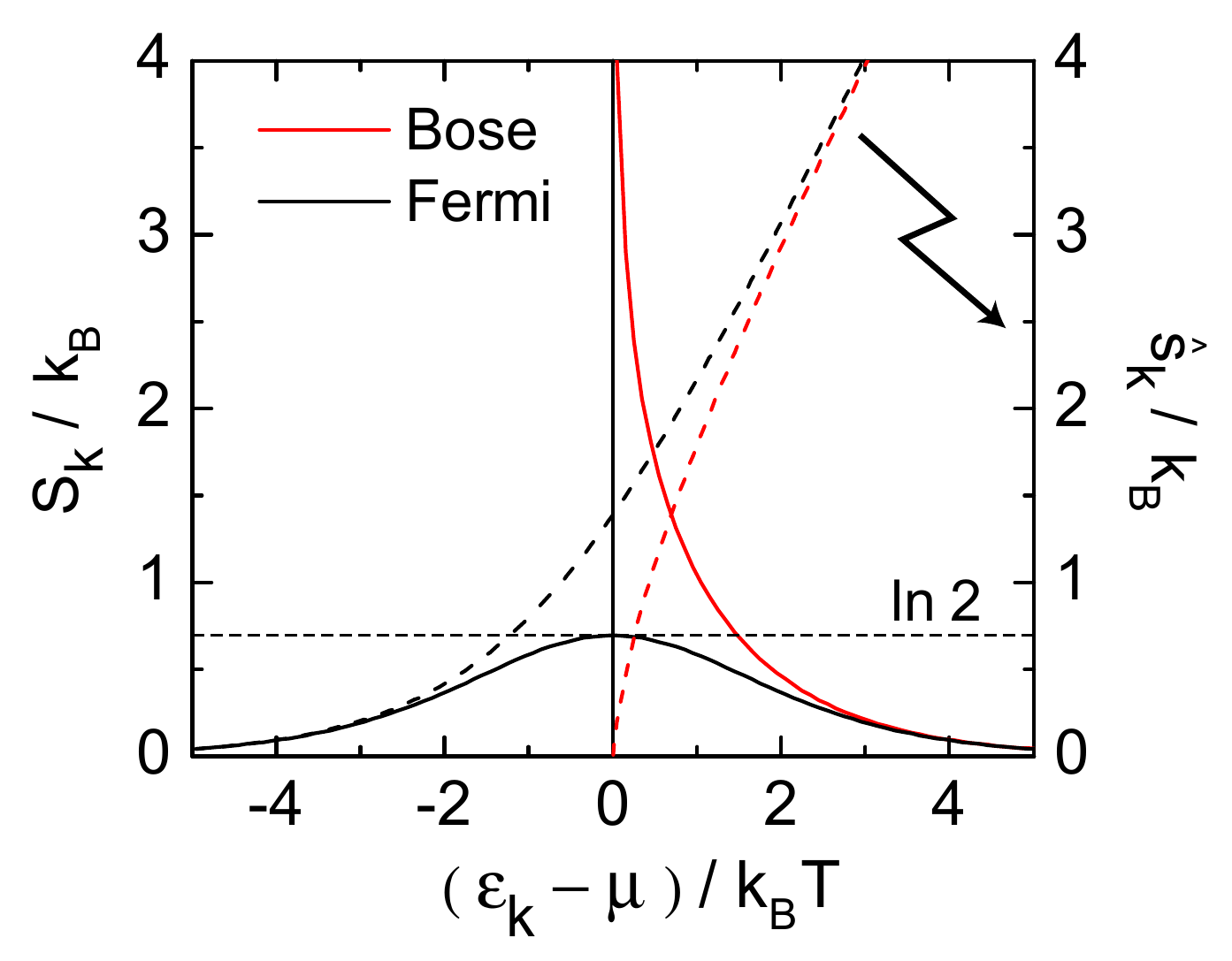}
\caption[]{ Entropy $S_k$ (solid lines) and entropy per particle $\shat_k$ (dashed lines) of elementary
Fermi- and Bose-systems with characteristic energy $\e(k)$.  }\label{fig:EntropieFB}
\end{figure}

The Landau-potential of an elementary Fermi- and Bose systems reads according to Eq.~\ref{gl_K_vs_ZG}
\begin{equation}\label{gl_K_el_FB}
K_{k}( T,V, \mubar) =\, \mp\,  k_BT\, \ln
\left\{ 1\,\pm\, \exp \left(- \frac{\varepsilon(k) -\mubar} {k_BT}\right) \right\}\;,
\end{equation}
where the upper sign hold for Fermi-, and the lower sign for Bose-systems, respectively.
Next we compute the entropy of the elementary  Fermi- and Bose-systems.
Using Eq.~\ref{gl_K_landau} and~\ref{gl_K_vs_ZG} one finds
\begin{eqnarray}\label{gl:S_Fermi}
S_k(T,V,\mubar)&=&-\frac{K_k-\big(\e(k)-\mubar\big) N_k}{T}\\\nonumber &=&\ \pm
\, k_B\,\left\{\ln\big(1\pm \exp(-Y_k)\big)
\pm\,\frac{Y_k}{\exp(Y_k)\pm1}\right\}\;,\end{eqnarray}with the abbreviation
$Y_k=\big(\e(k)-\mubar\big)/k_BT$. The upper sign refers to Fermi-
 and the lower sign to Bose-systems, respectively. For high energies $\e\gg\mubar$ the second term in Eq.~\ref{gl:S_Fermi} dominates, and the entropy per particle approaches (see Fig.~\ref{fig:EntropieFB}).
\begin{equation}\shat_k=\frac{\e(k)-\mubar}{T}\;.\end{equation}
 The same results are obtained from the thermodynamic derivative
$S_k(T,V,\mubar)=-\partial K_k(T,V,\mubar)/\partial T$.

To conclude this section, elementary Fermi- and Bose-systems are characterized by {\it two} equations of state, one for the particle number $N_\e(T,V,\mubar)$ (usually termed the {\it distribution function}), the other for the entropy $S_\e(T,V,\mubar)$. The first is ubiquitous in modern physics, while the latter is so far not much discussed in present literature. This is a pity, as it describes in a very concise form the caloric properties of these systems, which are the central building blocks of systems with indistinguishable particles. The reason for the lower prominence of $S_k$ is that the caloric properties of quantum gases are usually derived from the energy $E_\e(T,V,\mubar)=\e\cdot N_\e(T,V,\mubar)$. Since $\e(k)$ is not a variable, but a characteristic {\it constant} of the elementary subsystem labelled '$k$', $E_\e$ appears not to be independent from $N_\e$, and hence not as an independent equation of state. This is different for the entropy $S_\e(T,V,\mubar)$.

In usual terminology the elementary Fermi- and Bose-systems are called 'single particle states', as they are related to the solutions of the one-particle Sch\"{o}dinger equation. The Hilbert space of the many-particle system is represented by the tensor product of the spaces of the single particle systems. However, the  requirement of exchange symmetry resulting from the indistinguishability of identical particles eventually prevents the use of single-particle systems (whose Hilbert space is spanned by the solutions of the one-particle Schr\"{o}dinger equation) as independent building blocks of the many-body system. The exchange symmetry leads to quantum (Fermi- or Bose-like) correlations between the single particle systems, destroying their statistical independence. These correlations are much more conveniently taken into account in the framework of second quantization used in Eq.~\ref{gl_hamiltonian}. The Schr\"{o}dinger physics still enters in the form of the wave functions, which describe the spatial distribution of elementary Fermi- and Bose-systems. In other words, elementary Fermi- and Bose-systems are the excitation {\it modes} of the quantized matter field.
These are statistically independent, and hence, the thermodynamic potential of the gas is just the sum of the thermodynamic potentials over all elementary subsystems. The particle numbers $\{N_k\}$ and the entropies $\{S_k\}$ of the elementary subsystems are statistically independent random variables in the sense of thermodynamics, while the momenta of different particles are not independent, but subject to quantum correlations like the Pauli principle.

In the next step, we have to assemble the elementary Fermi- and Bose-systems to quantum gases. To do this, we assume that the $k$-vectors compatible with the boundary conditions lie dense enough in k-space, and convert the sum over all elementary systems in into an integral over the energies:\cite{footnote_k_e}
\begin{equation}\label{gl_sum_over_k}
\sum_{\vec{k}}\ =\ \frac{L^d}{(2\pi)^d}\int d^dk\ =\ L^d\int d\e\,g(\e)\;,
\end{equation}
where $g(\e)$ depends on the form of $\e(k)$ and the systems's dimensionality $d$; $L$ being the spatial extension of the system in each direction.
For spin 1/2 fermions, another factor of 2 has to be added or taken into account explicitly, if spin phenomena are studied. In usual terminology $g(\e)$ is called the density of (single particle) states (DoS), from our point of view, it is the density of elementary Fermi- and Bose-systems on the energy axis.

The {\it dispersion relation}  $\e(k)$ is (besides the Fermi- or Bose character) the main characteristic of the combined system, i.e., the specific quasi-particle gas under investigation. It is the only feature of our description that has a counterpart in classical physics (i.e., the function $E(\vec{P})$). The difference between the classical and the quantum point of view is the following:  rather than saying that the same (classically distinguishable) particle is accelerated by external forces or a scattering process with another particle, the quantum point of view is that particles in one elementary Fermi- or Bose-System are {\it annihilated}, while a (possibly different) number of particles is {\it generated} in other elementary Fermi-or Bose-System (of course under the constraint of the applicable conservation laws for the transitions). The dynamical variables are not anymore the positions and momenta of individual particles (which do not exist because of indistinguishability), but the particle number $N_{k}$, the entropy $S_{k}$, and the energy $E_{k}=\e(k)N_{k}$ of the elementary Fermi- and Bose-systems.

Using the integral representation of Eq.~\ref{gl_sum_over_k}, it is straightforward to compute the thermodynamic potential of the quantum gases from that of the elementary subsystems (see Eq.~\ref{gl_K_el_FB}):
\[K(T,V,\mubar)\ =  -k_BTV\int^\infty_0 d\e\, g(\e) K_\e(T,\mubar)\;,\]
where $V=L^d$.
 From the experimental point of view, it is more useful to compute the thermal
\begin{equation}\label{gl_Z_therm_FB_gase}
n(T,\mubar)\ =\  \int_0^\infty d\e\, g(\varepsilon) \cdot N_\varepsilon(T, \mubar)\;,
\end{equation}
and the caloric
\begin{equation}\label{gl_Z_kalor_FB_gase}
 e(T, \mubar)\ =\ \int_0^\infty d\varepsilon \, g(\varepsilon)
\cdot N_\varepsilon( T, \mubar) \cdot \varepsilon
\end{equation}
equation of state.
As a variant of the caloric EoS, the entropy density of an ideal gas is at all temperatures given by
\begin{equation}\label{gl_Z_S_FB_gase}
 s(T, \mubar)\ =\ \int_0^\infty d\varepsilon \, g(\varepsilon)
\cdot S_\varepsilon( T, \mubar) \;.
\end{equation}
Interestingly, all the entropy of the gas comes from the statistical mixture of states with different particle numbers within the same mode (or the same elementary subsystem). The combination of the elementary systems corresponding to different $k$-directions to the composite system does not enhance the entropy of the gas beyond the sum of the entropies of its elementary subsystems. This means that the entropy of a gas results exclusively from the statistical mixture of states with different particle numbers {\it within one} elementary subsystem! This will be important for the following section, where we will see that the reduced dimensionality of nanostructures leaves the properties of the individual elementary subsystem unchanged.

From the above equations of state all the thermodynamic susceptibilities entering the transport coefficients of the drift-diffusion model (see Sec.~\ref{sec:drift-diffusion})  can be computed in equilibrium, where all these systems share the same temperature and chemical potential.
For degenerate electrons, e.g. in three dimensions, one obtains in effective mass approximation the DoS
\[g(\e_F)\ =\ \frac{3}{2}\frac{n}{\e_F(n)}\ \propto n^{1/3}\;,\]
with the Fermi energy \[\e_F(n)=\mu(T=0,n)=\frac{\hbar^2(3\pi^2n)^{2/3}}{2\hat{m}}\;,\]
and one arrives at the well known thermodynamic susceptibilities:
\begin{eqnarray}\label{gl:particle capacity_degenerate e}
n^2\kappa_T&=&\frac{\partial n(T,\mu)}{\partial \mu}\,=\,g(\e_F)\\
\nonumber c_v(T,n)&=&\frac{\partial e(T,n)}{\partial T}\,=\, T\frac{\partial s(T,n)}{\partial T}\\\label{gl:entropy_degenerate e}
&=&\frac{\pi^2}{2}\,\frac{nk_B^2T}{\e_F(n)}\, =\, s(T,n)\\
\label{gl:ds_dn_degenerate e}
\frac{\partial s(T,n)}{\partial n}&=&\frac{\pi^2}{6}\,\frac{nk_B^2T}{\e_F(n)}\, =\, \frac{\shat(T,n)}{3}\;,
\end{eqnarray}
where $\kappa_T$ the isothermal compressibility, and $c_v$ the thermal capacitance per volume at constant particle density.

On the  other hand, for dilute gases, e.g., the electrons in semiconductors one obtains
\begin{eqnarray}\label{gl:particle capacity_dilute e}
n^2\kappa_T&=&\frac{\partial n(T,\mu)}{\partial \mu}\,=\,\frac{n}{k_BT}\\
\label{gl:entropy_dilute e}c_v&=&\frac{\partial e(T,n)}{\partial T}\,=\, T\frac{\partial s(T,n)}{\partial T}\,=\,\frac{3}{2}\,nk_B\\
\label{gl:s_dilute}
s(T,n)\,&=&\,\,nk_B\left\{\ln\left(\frac{jT^{3/2}}{{n}}\right)+\frac{5}{2}\right\}\;,\\
\label{gl:ds_dn_dilute e}
\frac{\partial s(T,n)}{\partial n}\,&=&\,\shat-k_B\,=\, k_B\left\{\ln\left(\frac{jT^{3/2}}{{n}}\right)+\frac{3}{2}\right\},
\end{eqnarray}
where $j=2\cdot(\hat{m}k_B/2\pi\hbar^2)^{3/2}$ is called the {\it chemical constant} of the electron gas (the prefactor 2 comes from spin). Eq.~\ref{gl:s_dilute} is also known as the Sackur-Tetrode equation. Eq.~\ref{gl:ds_dn_dilute e} provides a reasonable estimate for the thermopower of a doped (non-degenerate) semiconductor in the $T$-regime, where most dopants are ionized.\cite{note_TEP} Because of the small factor $k_BT_F/\e_F$ in Eq.~\ref{gl:ds_dn_degenerate e} the thermopower of metals is much smaller than that of semiconductors.

The fact that the state of macroscopic system with $\simeq 10^{23}$ internal degrees of freedom can be specified by only three independent variables (here $T$, $V$, and $\mu$) is a consequence of the thermal and chemical equilibrium between the elementary subsystems. The composite character of macroscopic systems becomes visible, if the equilibrium between the subsystems is disturbed, e.g., by a laser, which selectively increases the population of elementary subsystems at higher energy at the expense of those at lower energies. The equilibrium is restored by inelastic recombination processes.

The question is whether the elementary Fermi- and Bose systems are fundamentally relevant as separate entities, or are just a technicality, which allows an easy computation of the EoS. This question is equivalent to the question of why the grand canonical approach to statistical thermodynamics should be preferred to the micro-canonical and canonical ones. In equilibrium
and in the thermodynamic limit, where the $k$-space can be considered as a three-dimensional continuum, also the micro-canonical and the canonical approaches to thermodynamics work reasonably well. This changes when going beyond these restrictions as discussed in the next two sections. Historically, the characteristic statistical fluctuations of energy and particles numbers in these ensembles strongly biased the common view. Before the advent of quantum mechanics fluctuations were considered possible only for open systems.  Statistical fluctuations were a nuisance within classical physics, and their appearance within the canonical and grand canonical ensemble was viewed as an artifact, which ought to be removed by the thermodynamic limit, i.e., the limit $V,N\rightarrow\infty$ at constant particle density $n$. From this perspective, the micro-canonical approach is often considered as the most fundamental one, leading to the perception that thermodynamics as a whole works only within this limit. In quantum physics, however, statistical fluctuations constitute an unavoidable element of physics (see section~\ref{sec:discussion} for further discussion).


\section{Ballistic Quantum transport of particles}\label{sec_Q-transport}
In this section, it is shown that the elementary Fermi- and Bose-systems (which may look artificial at first sight) are very useful to understand transport properties in reduced dimensions, where the drift-diffusion model is entirely inapplicable. This is the regime of mesoscopic transport,\cite{ihn2010} where the size of the conducting object is smaller than the mean free path $\Lambda$.
More precisely, it is even sufficient, if the {\it inelastic} mean free path $\Lambda_{in}(T)$, corresponding to those scattering processes resulting in the dissipation of energy, is larger than the sample size.

Elastic scattering, i.e., scattering at constant energy,  cannot create entropy because any increase $\Delta S$ of entropy requires the amount $\Delta E=T\,\Delta S$ of energy, which by definition is not available. Instead, the elastic scattering modifies the underlying wave function, i.e., it changes the spatial distribution of the elementary Fermi- or Bose systems. The plane wave functions of ideal gases, or Bloch-wave functions  of crystals, respectively, are then replaced by complex interference patters, which are known as speckle patterns from laser physics.

At low temperatures $T\lesssim 1\;$K the inelastic mean free path $\Lambda_{in}(T)$ is typically in the micron regime, while in high mobility semiconductor also the {\it elastic} mean free path easily exceeds the micron range, and enables the study of truly ballistic transport. In this regime it is possible to experimentally realize some simple textbook quantum systems, where plane waves are scattered off tunable potential barriers.

The simplest case to consider is that of a quantum wire: a quantum wire can be realized for photons (wave guides or optical fibers comparable with the wave length of electromagnetic waves), for phonons (narrow beams with a diameter comparable to the wavelength of thermally excited phonons), and for electrons (semiconductor heterostructures or carbon nanotubes with a diameter comparable to the Fermi wavelength), to list a few examples. The common element in these examples is the fact that the set of allowed $k$-vectors forms not a three-dimensional continuum anymore, but consists of one, or a few one-dimensional
sub-continua, which propagate particles, entropy, and energy along the wire (say, in $x$-direction), while the transverse part of the wave function is discrete, resulting from the strong confinement of the system in the transverse direction. The one-dimensional sub-continua of elementary Fermi- or Bose-systems are also called {\it transport channels}. For simplicity, let us first assume that we have only one transport channel, as sketched in Fig.~\ref{fig:QPC}. Initially, we also assume that the quantum wire is perfectly transmitting (${\cal T}(\e)=1$); smaller transmission coefficients ${\cal T}(\e)<1$ are easy to take into account in the next step.

The transport of photons, phonons or electrons can then be viewed as a scattering problem: a beam of particles, emanating from two particle reservoirs connected to the left and right end of the wire, is either transmitted or reflected back. The elementary Fermi- or Bose-systems in the wire break up in two subsystems: {\it right movers} and {\it left movers}, which propagate particles, entropy, and energy with the dynamical velocity \begin{equation}\label{gl:dyn_velocity}
\vec{v}(k)=\frac{\partial \e(k)}{\hbar\,\partial k}\;,
 \end{equation}
 and are populated according to the temperature and the electrochemical potential of the left and right reservoir:\cite{note_reservoirs}
\[N_{\e;\,L,R}\ =\ N_\e(T_{L,R},\mubar_{L,R})\;.\]
In this approach the elementary Fermi- and Bose-systems can be visualized as {\it conveyor belts} for energy, entropy, particles, momentum, and spin, which transport these quantities ballistically, until an elastic, or inelastic scattering event occurs.

 If the two reservoirs differ in temperature or in (electro)\-chemical potential net currents of $E$, $S$, and $N$ will flow.
Despite being in a non-equilibrium state as a whole, the currents flowing through the wire of length $L$ are perfectly described by the thermodynamic properties of the two (left- and right-moving) subsystems. Following Landauer and B\"{u}ttiker\cite{landauer57,Buttiker86} we can write
\begin{eqnarray}\label{gl_j_1d_wire_perfekt}
I_N\ &=&\ \frac{1}{L}\left\{\sum_{k>0} N_{k,L}\,v(k)+\sum_{k<0}N_{k,R}\,v(k)\right\}\\\nonumber\,\\
\nonumber &=&\ \int\limits^\infty_{-\infty} d\e\;g(\e)\;v(\e)\;\big(N_\e(T_L,\mubar_L)-N_\e(T_R,\mubar_R)\big)\;,
\end{eqnarray}
where $\vec{v}\big(k\big)=-\vec{v}\big(-k\big)$. In the second step we have evaluated the sum in a continuum approximation using the one-dimensional DoS $g(\e)=(1/\pi)\cdot dk_x(\e)/d\e$ for propagating modes in one dimension.

If we plug this into Eq.~\ref{gl_j_1d_wire_perfekt} the energy dependence of the DoS and the velocity cancel and we obtain the surprisingly universal result
\begin{equation}
I_N\ =\  \frac{1}{\pi\hbar}\int\limits_{-\infty}^\infty d\e\;\big(N_\e(T_L,\mubar_L)-N_\e(T_R,\mubar_R)\big)\;,
\label{gl_j_1d_wire_perfekt_2}
\end{equation}
which is valid for both Fermi-  and Bose-systems and independent of the functional form of the dispersion relation $\e(k)$.
In solid state nanostructures, phonon and photon currents are usually not detected by measuring electric current or counting particles, but as a thermal (i.e. 'heat') current. We leave the treatment of thermal currents to the next section and specialize now to charged systems, in order to compute the electric conductance and the thermopower of quantum wires.

If the wire hosts several transport channels with $\e$-dependent transparencies ${\cal T}_n(\e)$, the charge current $I_Q=\hat{q}\,I_N$ assumes the more general form
\begin{equation}
I_Q\ =\  \frac{\hat{q}}{\pi\hbar}\sum_{n}\int\limits_{-\infty}^\infty d\e\;{\cal T}_n(\e)\,\big(N_\e(T_L,\mubar_L)-N_\e(T_R,\mubar_R)\big)\;.
\label{gl_j_1d_wire_real_2}
\end{equation}
Here, we limit ourselves now to the linear response regime, and assume that the wire is symmetrically biased, i.e.,
\begin{equation}\label{gl_QPC_bias}
N_{\e;\,L,R} =\ N_\e(T\pm\Delta T/2,\mu\pm \hat{q}U/2)\;,
\end{equation}
where $\Delta T$ and $U$ are the applied temperature and voltage bias, and the upper and lower sign refer to the left and right reservoir, respectively.
Then we can Taylor-expand the $N_{\e;\,L,R}$ of the reservoirs around the averages $(T_L+T_R)/2$ and $(\mubar_L+\mubar_R)/2$ and obtain in linear approximation ($k_B\Delta T,\hat{q}U\ll k_BT$):
\begin{eqnarray}\label{gl_delta_f}
N_\e(T_L,\mubar_L)&-&N_\e(T_R,\mubar_R)\ = \frac{\partial N(Y)}{\partial Y}\,\Delta Y\\
\nonumber&=&\ \frac{\partial N(Y)}{\partial Y}\underset{\Delta Y}{\underbrace{\frac{1}{k_BT}\left(\hat{q}U-\frac{\e-\mu}{T}\Delta T\right)}}\;
\end{eqnarray}
where $Y=(\e-\mu)/k_BT$, and
\begin{equation}\label{gl_dfdx}
\frac{\partial N(Y)}{\partial Y}\ =\ \frac{\exp(Y)}{\big(\exp(Y)\pm1\big)^2}\,.
\end{equation}
 The charge current then reads
\begin{equation}\label{gl_j_1d_wire_real_3}
I_Q=\frac{\hat{q}}{\pi\hbar}\sum_{n}\int\limits_0^\infty d\e\;{\cal T}_n(\e)\,\frac{\partial N(\e)}{\partial \e}\,\left(\hat{q}U-\frac{\e-\mu}{T}\Delta T\right)\;.
\end{equation}
In case of fermions, a Sommerfeld expansion of the integral in Eq.~\ref{gl_j_1d_wire_real_3}
leads to
\[I_Q\ =\ G\cdot U+G{\cal S}\cdot \Delta T\;,\]
where
\[G\ =\ G_0\cdot2\sum\limits_n{\cal T}_n(\mu)\;,\]
 (the factor 2 takes into account spin degeneracy) and
\begin{equation}\label{gl:cond_quantum}
G_0\ =\ \frac{\hat{q}^2}{h}\ \simeq\ 38.74\,\mu\textrm{S}\ \simeq
\ \frac{1}{25.8\,\textrm{k}\Omega}\end{equation}
is the universal {\it conductance quantum}. This is a seminal result of mesoscopic physics found experimentally first in gate-defined quantum point contacts.\cite{vwees88,wharam88}

The thermoelectric counterpart of the conductance quantization, i.e., the Seebeck coefficient of a quantum wire, or quantum point contact is given by
\begin{equation}\label{gl:mott_ballistic}{\cal S}\ =\ \frac{\pi^2}{3}\frac{k_B^2}{\qhat h}\cdot 2\sum\limits_n\,\left.\frac{d\big(\ln {\cal T}_n(\e)\big)}{d\e}\right|_{\e=\mu}\;.
\end{equation}
Also this result has been experimentally confirmed first in quantum point contacts.\cite{molenkamp91}

\section{Ballistic Quantum transport of entropy}\label{sec:S-transport}
The first efforts to transfer the ideas of ballistic electron transport to thermal transport originate from Imry and Sivan\cite{sivan86} and Butcher.\cite{butcher90}
To describe the thermal transport through quantum wires, we can write down an expression  for the entropy current that is the thermal analogue of Eq.~\ref{gl_j_1d_wire_perfekt_2}, but contains the entropy $S_{k;\,L,R}$ (Eq.~\ref{gl:S_Fermi}) of the elementary Fermi- or Bose-systems rather than their particle numbers $N_{\e;L,R}$. Assuming that the entropy propagates in each elementary Fermi-, or Bose-system at the same velocity $\vec{v}(k)$ as the particles, the entropy current reads:
\begin{equation}
 I_S\ =\  \frac{1}{\pi\hbar}\int\limits_{-\infty}^\infty d\e\;\big(S_\e(T_L,\mubar_L)-S_\e(T_R,\mubar_R)\big)\;.
\label{gl_jS_1d_wire_perfekt_2}
\end{equation}
For the same symmetric bias (see Eq.~\ref{gl_QPC_bias}) one finds  in linear approximation
\begin{eqnarray}\label{gl_delta_S}
S_\e(T_L,\mubar_L)&-&S_\e(T_R,\mubar_R)\ = \frac{\partial S(Y)}{\partial Y}\,\Delta Y\\
\nonumber&=&\ \frac{\partial S(Y)}{\partial Y}\frac{1}{k_BT}\left(\hat{q}U-\frac{\e-\mu}{T}\Delta T\right)\;,
\end{eqnarray}
where again $Y=(\e-\mu)/k_BT$, and
\begin{equation}\label{gl_dSdx}
\frac{\partial S(Y)}{\partial Y}\ =\ k_BY\cdot\frac{\exp(Y)}{\big(\exp(Y)\pm1\big)^2}\;.
\end{equation}
The upper sign holds for fermions, and the lower for bosons.
Interestingly, this result differs from Eq.~\ref{gl_dfdx} only by the extra factor $k_BY$.
For the entropy current we then find in linear response
\begin{equation}\label{gl_jS_1d_wire_real_3}
I_S=\frac{1}{\pi\hbar}\sum\limits_n\int\limits_0^\infty d\e\;{\cal T}_n(\e)\,\frac{\e-\mu}{T}\frac{\partial N(\e)}{\partial \e}\,\left(\hat{q}U-\frac{\e-\mu}{T}\Delta T\right).
\end{equation}
The first term ($\propto \Delta\mubar$) in this equation is driven by the voltage bias, and constitutes the ballistic analogue to the Peltier current in Eq.~\ref{gl_j_S}. The second term ($\propto \Delta T$)  describes the entropy current driven by the $T$-difference.

The same result is obtained, if one extends the derivation of $I_N$ to $I_E$ and computes $I_S$ via Eq.~\ref{gl_GFF_J}:
\begin{equation}\label{gl_jS_1d_wire_real_4}
\displaystyle I_S\ =\ \frac{1}{T}\Big(I_E-\mubar I_N\Big)\ =\ \frac{1}{T}\Big\{\Pi\cdot\,U\ +\ {\cal L}\cdot\Delta T\Big\}
\end{equation}
Note that the identification of $I_S$ with $(I_E-\mubar I_N)/T$ holds only in the linear response regime.\cite{note_nonlinear}

For fermions, we can again evaluate the integrals in Eq.~\ref{gl_jS_1d_wire_real_3} within the Sommerfeld approximation, and obtain for the Peltier coefficient
\[\Pi\ =\ T\cdot \frac{2L_0}{\qhat} \,\left.\sum\limits_n\frac{d\big(\ln {\cal T}_n(\e)\big)}{d\e}\right|_{\e=\mu}\;,\]
and for the thermal conductance
\[{\cal L}\ =\ T\cdot 2L_0\,\sum\limits_n\left.{\cal T}_n(\mu)\right.\;,\]
where the prefactors $2$ accounts again for spin-degeneracy.
The quantity $L_0$ is the {\it entropy conductance quantum} corresponding to
\begin{equation}\label{gl_term_cond_quantum}
L_0\ =\ \frac{\pi^2}{3}\,\frac{k_B^2}{h}\ =\  0.9456~\text{pW/K}^2
\;,
\end{equation}
implying that {\it in ballistic quantum wires the entropy conductance $L={\cal L}/T$ is quantized in units of $L_0$}. Compared to the quantum of electric conductance, $\hat{q}^2$ is replaced by $(\pi k_B)^2/3$.

As discussed in the context of the drift-diffusion model in Sec.~\ref{sec:drift-diffusion} the Seebeck- and Peltier-coefficients are connected by the Kelvin-Onsager relation
\begin{equation}\label{gl:kelvin_ballistic}
\Pi\ =\ T\cdot{\cal S}\;.
\end{equation}
 This results from the Maxwell-relation
\begin{equation}\label{gl:maxwell_S_e_N_e}
\frac{\partial N_\e(T,\mubar)}{\partial T}\ =\ \frac{\partial S_\e(T,\mubar)}{\partial \mubar}\ =\ -\frac{\exp(Y)}{T\big(\exp(Y)\pm1\big)^2}\;,
\end{equation}
 between the derivatives of the two EoS (Eqs.~\ref{gl_Z_therm_FB_gase} and \ref{gl_Z_S_FB_gase}), and the fact that the transmission coefficients ${\cal T}(\e)$ determine all transport quantities.\cite{onsager2}

These results are again very general, as they depend only on the energy dependence of the transmission coefficients, but neither on the dispersion relation nor on the particle statistics. After some more qualitative experiments\cite{molenkamp92} a quantitative experimental investigation of the thermal conductance of quantum point contacts has been performed only recently, exploiting the thermopower of quantum point contacts for local thermometry.\cite{chiatti2006}
Very recently, also chiral thermal transport in the integer quantum Hall regime has been demonstrated.\cite{granger2009}

The so far most relevant cases for bosons deal with phonons and photons. In this case we can set $\mu=0$. The evaluation of the corresponding Bose-integral results in a quantized thermal conductance with the very same entropy conductance quantum $L_0$ as in the case of fermions.\cite{rego98_99,blencowe99} The case of the quantized entropy conductance by phonons was first addressed by the beautiful experiments by Schwab {\it et al.}\cite{schwab2000} A few years later, Meschke {\it et al.} considered the case of microwave photons.\cite{meschke} The latter case is of particular importance for instrumentation in mesoscopic physics. First, it explains why a careful filtering of the measurement leads at low temperatures is required: in a cryogenic setup the wires bring down energy and entropy not only via electronic and phononic thermal conduction, but also via thermal photons. These photons may not carry enough energy to heat macroscopic objects like the thermometer, but they can still induce jumps of charge carriers in electronic traps, or release single electrons from a quantum dot.
Second, the photon case is interesting, because the techniques of microwave engineering
provide possibilities to manipulate the photon thermal transport at the mesoscopic scale.\cite{niskanen2007}

%

\section{Connection to diffusive transport}\label{sec:boltzmann_eq}
\begin{figure}[t]
\includegraphics[width=66mm]{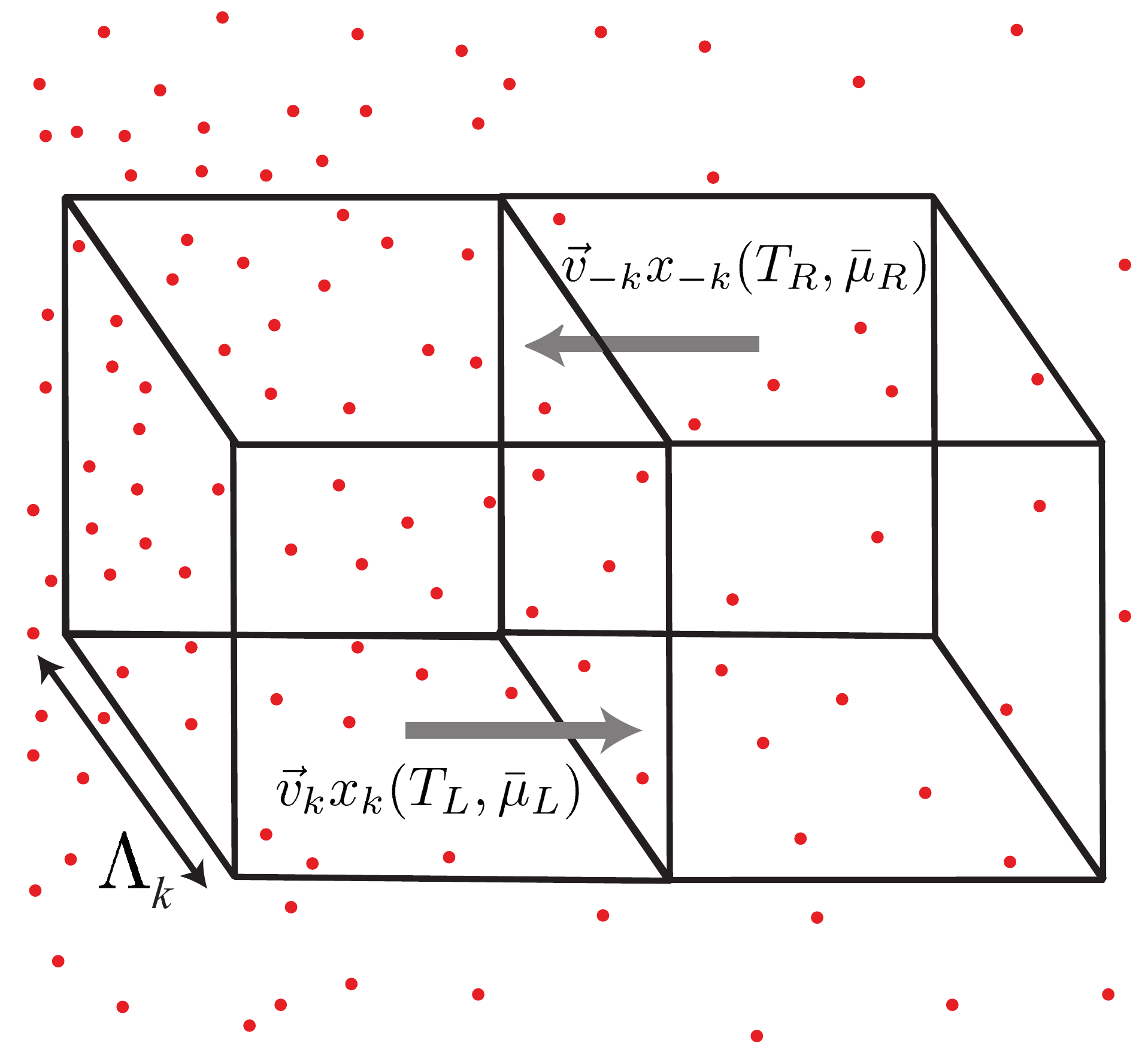}
\caption{\label{fig:diffusion_2}
The left- (right-) propagating elementary Fermi- or Bose-systems labelled '$\vec{k}$' emanating from the right (left) volume element of size $\Lambda_k$ contribute an amount $v_kx_k(T_{L,R},\mubar_{L,R})$ to the total $X$-current density $\vec{j}_X$.
}
\end{figure}

The concept of elementary Fermi- and Bose-systems can also be used to analyze the case of diffusive transport more accurately than in the drift-diffusion model. To do this, we consider in analogy with section~\ref{sec:drift-diffusion} two adjacent volume elements of linear dimension $\Lambda=|\vec{v}|\cdot \tau$, which are illustrated in Fig.~\ref{fig:diffusion_2}, and assume that an inelastic scattering mechanism exists, which establishes local thermal and electro-chemical equilibrium on the scale of $\Lambda$. Volume elements of this size represent the smallest possible units, which can be assumed to be in local equilibrium. For a given $T$- or $\mubar$-gradient, two adjacent cubes exhibit the minimal $T$- or $\mubar$-difference possible.

Between collisions the elementary Fermi- and Bose-systems propagate energy, entropy and particles ballistically, as they do in the one-dimensional ballistic quantum wires discussed in the preceding sections. To account for the three dimensions we have to average over the different $k$-directions, in order to determine the current density of any balanceable quantity $X$ in a given direction.

Similar to section~\ref{sec:drift-diffusion}, we write the $X$-current density through the interface between the  two cubes as
the difference
\begin{equation}
\vec{j}_X\ =\ \sum\limits_k\vec{v}(k)\cdot\Big(x_{k}(T_L,\mubar_L)-x_{k}(T_R,\mubar_R)\Big)\;
\end{equation}
of left and right propagating currents, where $x_k$ is the contribution of each elementary subsystem to the total $x$-density depending on the values of $T_{L,R}$ and $\mubar_{L,R}$ within two adjacent cubes of size $\lambda_k^3$ (see Fig.~\ref{fig:diffusion_2}).
The similarity of this expression to those used in ballistic transport is not accidental, but results from the fact that the propagation of any balanceable quantity $X$ is ballistic over distances smaller than the mean free path $\Lambda_k$.
In the examples considered here $x_k$ is the contribution $n_k=N_k(T,\mubar)/V$ or $s_k=S_k(T,\mubar)/V$ of an elementary Fermi- or Bose-system  with the wave vector $k$ to the particle density, or the entropy density, respectively.

In contrast to the more elementary treatment of the diffusive limit in section~\ref{sec:drift-diffusion} we now take into account that $\vec{\Lambda}_k=\vec{v}(k)\cdot\tau_k$ depends via both $\tau_k$ (see Ref.~\onlinecite{note_rel_time_approx}) and the velocity $\vec{v}(k)$ of propagation (see Eq.~\ref{gl:dyn_velocity}) between collisions depend on $\e$, or more precisely on $k$.
In linear approximation, the difference $\Delta x_k$ reads in analogy to the preceding sections~\ref{sec_Q-transport} and~\ref{sec:S-transport}
\begin{equation}\label{gl:Delta_x_k}
x_{k,L}-x_{k,R}\ =\ \frac{\partial x_k(Y)}{\partial Y}\,\Delta{Y}\;,
\end{equation}
where again $Y=(\e-\mu)/k_BT$, and $\Delta Y$ is the variation of $Y$ over the distance of ballistic propagation between scattering events.
Similar to Eq.~\ref{gl_delta_f} $\Delta Y$ is given by
\begin{equation}
\Delta Y_k\ =\ \frac{1}{k_BT}\,\left(\nabla \mubar-\frac{\e-\mu}{T}\;\nabla T\right)\cdot\underset{\vec{\Lambda}_k}{\underbrace{\vec{v}(k)\tau(k)}}\;.
\end{equation}
Here, $|\vec{\Lambda}_k|$ is the mean free path associated with the scattering of quasiparticles in the elementary subsystem with wave vector $k$.

   Plugging $\Delta Y_k$ into Eq.~\ref{gl:Delta_x_k}  we obtain in the continuum limit
\begin{equation}\label{gl:X_transport}
\vec{j}_X\ =\ \int\limits_0^\infty d\e\;g(\e)\boldsymbol{D}(\e)\frac{\partial x(\e)}{\partial \e}\left(\nabla \mubar-\frac{\e-\mu}{T}\;\nabla T\right)\;,
\end{equation}
where the diffusivity tensor
\begin{equation}\label{gl:D_k}
\boldsymbol{D}(\e)\ =\ \iint\limits d\theta d\varphi\;\vec{v}(\e,\theta,\varphi)\otimes\vec{v}(\e,\theta,\varphi)\cdot\tau(\e,\theta,\varphi)
\end{equation}
is the energy-dependent tensorial equivalent of the diffusion constant (averaged over the angles $\theta$ and $\varphi$ on a sheet of constant $\e(k)$), which takes into account
the shape and angle anisotropy of the dispersion relation $\e(k)$ and the scattering time $\tau(k)$.\cite{ashcroft}
For an isotropic $\e(k)$ and $\tau(k)$ Eq.~\ref{gl:D_k} reduces to Eq.~\ref{gl:diff_const}.

Equation~\ref{gl:X_transport} constitutes the linearized transport equation for any balanceable quantity $X$, and is in perfect agreement with the result from the solutions of the Boltzmann equation in relaxation-time approximation in textbooks.\cite{ashcroft} However, it is remarkable that its derivation here is based not on the classical concepts of trajectories, but on the sole assumption of the existence of a scattering mechanism ensuring that the elementary Fermi- and Bose systems propagate (on average) ballistically from opposite faces through a cube of size $\Lambda^3_k$ have values of $x_k(T,\mubar)$ according to the local values of $T(\vec{r})$, $\mubar(\vec{r})$, and $T(\vec{r}+\vec{\Lambda}_k)$, $\mubar(\vec{r}+\vec{\Lambda}_k)$, respectively.

The evaluation of Eq.~\ref{gl:X_transport} for the particle and the entropy current reproduces the Drude formulas for the electric (Eq.~\ref{gl_Einstein}) and the thermal (Eq.~\ref{gl:lambda}) conductivity, as well as the Wiedemann-Franz law. The evaluation of the thermopower and the Peltier coefficient within the Sommerfeld expansion leads to a slightly modified result, when comparing to the drift-diffusion model (see Eq.~\ref{gl_j_Q2}). While the thermodynamic derivative $\partial n(T,\mu)/\partial T$ entering Eq.~\ref{gl_seebeck} involves only the derivative $dg(\e)/d\e$ (because the diffusion constant $D$ is assumed to be independent of $\e$), the evaluation of  Eq.~\ref{gl:X_transport} contains the derivative $d\big(g(\e)\boldsymbol{D}(\e)\big)/d\e$. The thermopower is then given by the Mott-formula
\begin{equation}\label{gl:mott_diffusive}{\cal S}_\text{mott}\ =\ \frac{\pi^2}{3}\frac{k_B^2T}{\hat{q}}\cdot \,\left.\frac{d\big(\ln [g(\e){\boldsymbol D}(\e)]\big)}{d\e}\right|_{\e=\mu}\;,
\end{equation}
which is the diffusive analog of Eq.~\ref{gl:mott_ballistic}. If both $g(\e)\propto\e^\alpha$ and $\boldsymbol{D}(\e)\propto \e^\beta$ obey a power law, the logarithmic derivative in Eq.~\ref{gl:mott_diffusive} assumes the value
\[\left.\frac{d\big(\ln g(\e){\boldsymbol D}(\e)\big)}{d\e}\right|_{\e=\mu}\ =\ \frac{\alpha+\beta}{\e_F}\;,\]
as opposed to the result $\alpha/\e_F$ obtained within the drift-diffusion model. For free electrons in three dimensions we have $\alpha=1/2$ and $\beta=1$, if we assume the scattering time $\tau$ to be independent of $\e$. In this simplest approximation, the result of Eq.~\ref{gl:mott_diffusive} is a factor of three larger than Eq.~\ref{gl:ds_dn_degenerate e}, and we have the curious relation
\begin{equation}\label{gl:seebeck_vs_shat}
{\cal S}_\text{mott}\ =\ \frac{\shat(T,n)}{\hat{q}}\;.
\end{equation}
This result is sometimes exploited to estimate the molar entropy of strongly correlated Fermi systems from measurements of the thermopower. In doing so, one has of course to be aware, that Eq.~\ref{gl:seebeck_vs_shat} heavily relies on a quadratic $\e(k)$ and the energy-independence of $\tau$. It is certainly much safer to extract $\shat$ from measurements of the molar heat capacity.

So far we have restricted our consideration to an isolated degenerate Fermi gas. A more accurate modelling has to take into account that a solid usually contains several different Fermi- or Bose-gases. Besides the electrons, there are phonons, and (in magnetic solids) also magnons, or localized magnetic moments. The interaction between the different quasiparticle systems provides usually scattering mechanisms with an energy-dependent scattering time $\tau(\e)$, which may deviate from a power law, or at least, affect the value of $\beta$. In addition, there can be drag phenomena, like electron-phonon, and electron-magnon drag, which can substantially complicate the behavior of ${\cal S}$.\cite{barnard}

The mean free path $\Lambda_k$ (given by Eq.~\ref{gl_lambda}) can be calculated from the solution of the quantum mechanical scattering problem of the relevant particles. It is quite ingenious that the model describes the {\it irreversible} process of conduction, without the need to explain how entropy is actually generated in the scattering process. As we will discuss in section~\ref{irreversibility}, the scattering cross section is calculated within standard Hamiltonian quantum theory, in which the time evolution is always reversible. Nevertheless, as the scattering enters only via a scattering probability (and not via a probability amplitude) in Eq.~\ref{gl:X_transport}, it is assumed that the phase coherence in the scattering process is {\it quenched}. This looks like a rather arbitrary truncation of the Hamiltonian time evolution that has some resemblance with the quantum mechanical measurement process. It is this (here manually imposed) quenching of the phase coherence, which gives rise to the 'classical' character of the Boltzmann-like transport theory.
In the next section it is discussed, how phase coherence affects the transport. It turns out experimentally, that it is the inelastic scattering processes, which account phenomenologically for the loss of phase coherence.

\section{Discussion}\label{sec:discussion}
\subsection{Applicability of Thermodynamics}
It is a widely spread opinion that the concepts of thermodynamics are applicable only in equilibrium, and for large systems with many microscopic degrees of freedom. The limit of large systems is also called the thermodynamic limit, where $N,\,V\rightarrow\infty$, and $n=N/V=const$. These restrictions  result from the custom to {\it define} entropy by Boltzmann's famous formula
\begin{equation}
S\ =\ k_B\,\ln \Omega(E,V,N)\;,
\end{equation}
where $\Omega$ is the number of microstates accessible to $N$ particles in a volume $V$ with a given value of the total energy $E$. This strategy (when properly combined with the principle of indistinguishability) indeed allows to calculate the function $S(E,V,N)$, which is equivalent to $E(S,V,N)$, and hence constitutes a starting point for equilibrium thermodynamics in entropy representation.
However, in so-called {\it open} systems, which exchange energy and particles with the environment, the number $\Omega$ of microstates cannot be defined anymore, and hence it seems that thermodynamics is inapplicable to transport situations.

On the other hand, the presentation of the preceding sections shows that thermodynamics {\it can} well be applied to a large variety of non-equilibrium and transport situations, {\it provided that  a proper decomposition into simpler sub-systems is chosen}. Each of these elementary sub-systems is at least approximately in thermodynamic equilibrium with one of the reservoirs, while the thermal and chemical equilibrium with its fellow sub-systems propagating in the opposite direction can be disturbed.

\begin{figure}[t]
\centering
\includegraphics[width=85mm]{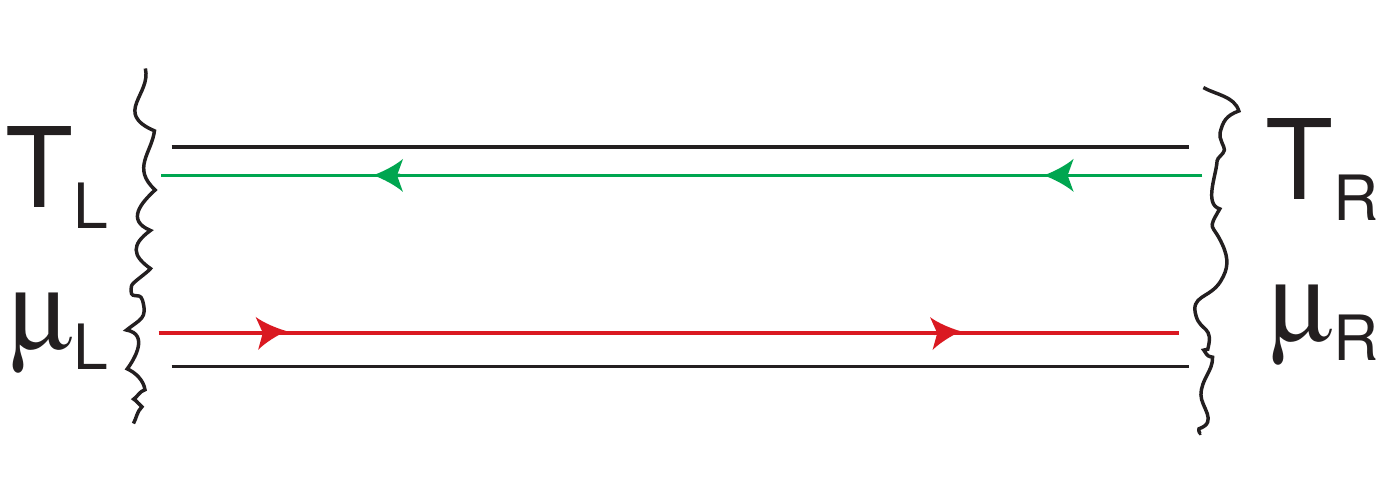}
\caption[]{Schematic of single channel quantum wire connected to two reservoirs for energy, entropy, and particles. The one-dimensional character of the transport can be realized by the formation of edge states in a quantizing magnetic field, i.e., in the quantum Hall regime. In this case the magnetic  provides also a spatial separation between left- and right-movers. The elementary Fermi-or Bose systems are charged with energy, entropy, and particles via the left (red) and the right (green) reservoir, respectively. They are in thermodynamic equilibrium with their source reservoirs, but not with each other.}\label{fig:Qwire}
\end{figure}

To be more specific, in a ballistic quantum wire such as the one-dimensional subbands formed in high mobility two-dimensional electron systems (see Fig.~\ref{fig:Qwire}), the set of right- (left-)moving elementary Fermi, or Bose systems are in equilibrium with each other and the left (right) reservoir, but the equilibrium between left and right movers is disturbed by the applied bias voltage, or the temperature difference, respectively. The same holds locally in a macroscopic piece of matter subjected to a $T$- or $\mubar$-gradient: also here left- and right moving elementary Fermi- and Bose-systems are out of equilibrium, but the amount of entropy and particles stored in the elementary Fermi- or Bose-systems {\it propagating in the same direction from a certain point in space} is to a very good approximation given by the local values of the temperature and (electro)-chemical potential.

This explains, why the current densities of entropy and particles are determined by the diffusion coefficient and the local (equilibrium) values of the derivatives of densities $n_k(T,\mubar)$ and $s_k(T,\mubar)$, respectively.

In this sense the thermodynamic concepts retain their relevance, irrespective of a reduced dimensionality or the absence of global equilibrium. This is also intuitively clear, as we don't question the validity of the concept of water temperature based on the undisputable absence of thermal equilibrium between the Mediterranean and the polar sea. That the same concepts remain valid in quantum wires with perfect transmission down to the atomic scale, e.g., in highly transparent atomic point contacts, appears more surprising, but has been demonstrated in a variety of beautiful experiments.\cite{elke}

 It has to be pointed out that a large body of literature considers the Hamiltonian dynamics of small quantum systems coupled to reservoirs, striving for a modelling of nanoscale thermodynamic processes (see, e.g., Ref.~\onlinecite{nieuwenhuizen_mahler} and the references therein), including the transfer of work and heat between nanosystems and the reservoirs.
In contrast the present work is focused on simple transport phenomena, and tries to elucidate the conceptual foundation, on which the standard approach to both ballistic and diffusive transport is so successful.

\subsection{Breakdown of Local Equilibrium}
The Landauer-B\"{u}ttiker approach to transport restricts itself to the particular case of a nanoscale constriction between macroscopic reservoirs, in which the \textit{thermodynamics of the elementary Fermi- or Bose-systems} in the constriction is governed by the reservoirs, and can be separated from the \textit{transmission properties} of the constriction expressed by the set $\{{\cal T}_n(\e)\}$ of transmission coefficients.\cite{footnote_L0} This separation holds best in the linear regime, and as long as energy and particle number of the elementary subsystems differing in the characteristic energy $\e$ are statistically independent, because no inelastic scattering induces an exchange of energy, entropy and particles between them. In this way the population of the reservoirs, which are by definition in thermal equilibrium, is transferred to the elementary subsystems. These can be considered in thermodynamic equilibrium with one of the reservoirs, while there is no equilibrium between elementary subsystems charged by {\it different} reservoirs.

Under these conditions, the propagation of energy, entropy and quasiparticles within the constriction is governed by reversible Hamiltonian dynamics, while the conceptionally difficult irreversible equilibration of the injected quasiparticles within the reservoirs {\it is irrelevant for the transport properties of the constriction}! This explains the success of the Hamiltonian dynamics in the description of this category of transport processes. It is interesting that this success can (within the semi-classical approximation) also be transferred to the macroscopic case, where reservoirs are absent, and a sufficiently strong inelastic scattering ensures local thermodynamic equilibrium, if one averages over volumes larger than the mean free path (see Sec.~\ref{sec:boltzmann_eq}).

\begin{figure}[t]
\centering
\includegraphics[width=85mm]{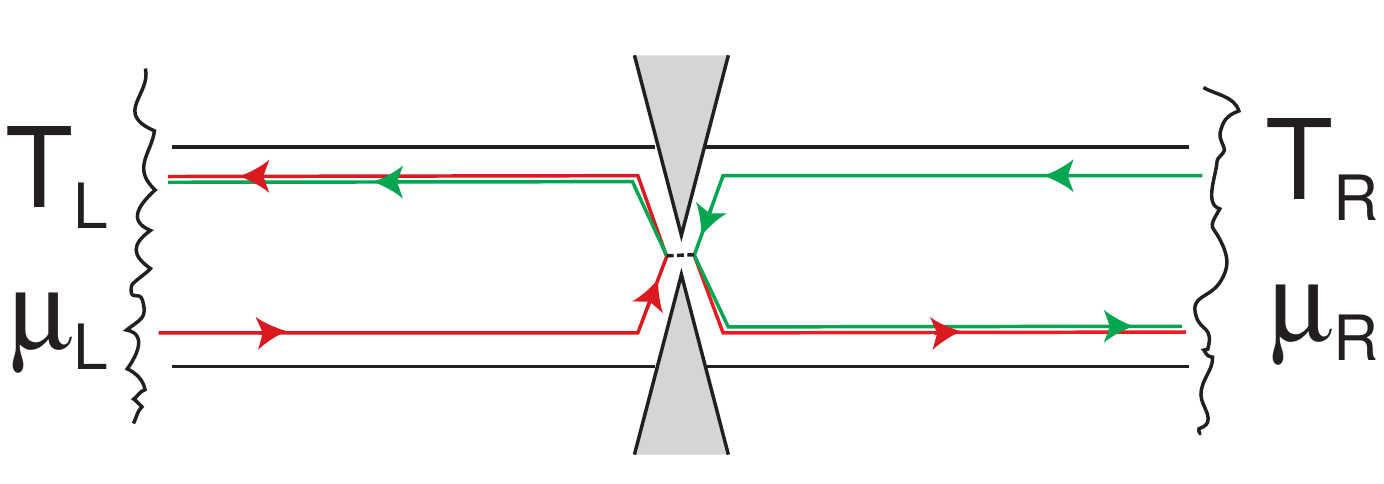}
\caption[]{The same quantum wire as in Fig.~\ref{fig:Qwire}, but interrupted by a quantum point contact (QPC) with transmission coefficient ${\cal T}(\e)$. The scattering states of the particles emanating from the left (red) and the right (green) reservoir, respectively, have to be considered as {\it one}  elementary Fermi- or Bose-systems, which cannot be further decomposed into subsystems.}\label{fig:QPC}
\end{figure}

If elastic scattering occurs within the phase coherent region, the elementary subsystems labelled $\{k\}$ in the absence of scattering have to be replaced by more complicated ones, which in usual terminology are called {\it scattering states}, and consist of the incoming and the two (or more) outgoing waves. Such a complex wave pattern still forms {\it one} elementary Fermi- or Bose system that cannot be decomposed further. In particular, it is impossible to provide a local thermodynamic description of the quantum wires left and right of the QPC in Fig.~\ref{fig:QPC}. The QPC partitions the flux of energy, entropy and quasi-particles emanating from each reservoir according to the transmission coefficient ${\cal T}(\e)$ into the two corresponding outgoing fluxes. As the particles emanating from different reservoirs are {\it incoherent}, and thus statistically independent, their average particle numbers in the elementary Fermi-and Bose systems flowing into the right reservoir simply add up according to
\begin{equation}\label{gl:sup_N1}
N_{k}^\text{neq}\ =\ {\cal T}(\e)\cdot N_{k}^\text{eq}(T_L,\mubar_L)+(1-{\cal T}(\e))\cdot N_{-k}^\text{eq}(T_R,\mubar_R)\;,\end{equation}
 while we have for the left reservoir
\begin{equation}\label{gl:sup_N2}
N_{-k}^\text{neq}\ \ =\ (1-{\cal T}(\e))\cdot N_{k}^\text{eq}(T_L,\mubar_L)+{\cal T}(\e)\cdot N_{-k}^\text{eq}(T_R,\mubar_R)\;.\end{equation}
Here $N_{k}^\text{eq}$ denotes the equilibrium particle numbers in the reservoirs. According to Eq.~\ref{gl_j_1d_wire_perfekt_2} the non-equilibrium particle numbers  $N_{k}^\text{neq}$ determine the current.  Analogous expressions hold for $S_k$, $E_k$, and all other balanceable quantities of the system.

Being a superposition of equilibrium particle numbers, and entropies with different $\{T,\mubar\}$, the particle numbers $N_k$ in the outgoing branches  do not satisfy Eq.~\ref{gl_bosefkt} anymore. This signals the break-down of local equilibrium.
Similarly, the entropies ${\cal T}(\e)S_k$ and $\big(1-{\cal T}(\e)\big)S_k$ pro\-pagated by the outgoing partial waves after the scatterer are {\it lower} than the entropy of an equilibrium mode with the same temperature of the left reservoir (see Eq.~\ref{gl:S_Fermi}) and particle numbers ${\cal T}(\e)N_k$ and $\big(1-{\cal T}(\e)\big)N_k$, respectively.

One may be tempted to consider a single quantum wire between the scattering region and one reservoir as a 'system' on its own, and ask for its thermodynamic quantities, e.g., its particle number or entropy.  This object, however, is certainly not an independent sub-system in the sense of thermodynamics, because its particle numbers and other physical quantities are correlated with those of the other quantum wires connected to the constriction. The correlation of the particle numbers can be experimentally accessed, e.g., as an anti-correlation between the particle currents in the two reservoirs.\cite{electronic_HBT} Thus it has to be kept in mind that the transport modes, i.e., the set of elementary Fermi- and Bose-systems hosted by the scattering region and the quantum wires cannot be decomposed in smaller subsystems as long as their phase coherence is not destroyed by inelastic processes.

A particularly nice experimental demonstration of such out-of-equilibrium physics was recently obtained using the one-dimensional edge channels in the quantum Hall regime.\cite{altimiras10,leSueur10} There it was possible to measure the $\e$-dependence of the particle numbers $N_\e$ at different distances (0.8-30~$\mu$m) from a quantum point contact. Close to the quantum point contact a double step shape of $N_\e$ was observed, which resulted from the superposition of the transmitted and reflected electrons emanating from of reservoirs at different electrochemical potential (see Eqs.~\ref{gl:sup_N1} and \ref{gl:sup_N2}). At larger distances inelastic scattering processes restored the local thermal and chemical equilibrium, and $N_\e$ behaved Fermi-like. The measured local temperature was higher than that in the reservoirs, but lower than expected from energy conservation arguments.
One possibility to explain the apparent loss of energy are additional {\it neutral modes}, which are predicted to exist, when two (spin-degenerated) edge channels are present.\cite{levkivski08}

A breakdown of local equilibrium can also happen in the diffusive regime. As it was shown in a series of experiments on diffusive wires of length $L$ in the $\mu$m-regime this is possible at low temperatures, if the average diffusion time $\tau_D=L^2/D$ through the wire is shorter than the time required for energy relaxation in the wire.\cite{pot97} This means the energy relaxation occurs predominantly by inelastic scattering in the two macroscopic reservoirs serving as contacts for the wire. If the energy relaxation in the wire is entirely negligible, no particle exchange occurs between elementary Fermi-systems of different characteristic energy $\e$ -- i.e., elastic scattering processes dominate in the wire. The effect of the elastic scattering is an efficient randomization\cite{note10} of the momentum distribution in the wire, which corresponds to intense exchange of particles (and entropy) between elementary Fermi-systems with the same $\e$, but differing in $k$-direction.\cite{note11}
In absence of inelastic scattering the elementary Fermi-systems with the same $\e$ are not in local equilibrium, because $N_\e$ and $S_\e$ at any point in the wire have contributions from both reservoirs.

To quantify these considerations, it is simplest to consider the diffusion equation for the particle density $n_\e(r)$ in elementary Fermi-systems with the same $\e$ ($r$ is the position along the wire) with a source term $\Sigma_{N_\e}$ describing the exchange of particles with elementary Fermi-systems with other $\e$. The diffusion equation results from the combination of the continuity equation Eq.~\ref{continuity}  with Fick's law (Eq.~\ref{gl_DDmodel_N}) for the particle densities $n_\e(r)$ in the composite sub-system containing all elementary Fermi systems with the same characteristic energy $\e$:
\begin{equation}\label{gl:diffusion}
\frac{\partial n_\e(\vec{r},t)}{\partial t}+\mathrm{divgrad}\,n_\e(\vec{r},t)\ =\ \Sigma_{N_\e}\;.
\end{equation}
The source term $\Sigma_{N,\e}$ is called the {\it scattering integral}, and consists of a sum over all possible transitions between one elementary Fermi-system and the others with the quantum-mechanical transition probabilities $W_{\e,\e'}$.\cite{pot97} For isotropic scattering only the energy dependence of the $W_{\e,\e'}$ is relevant, and it reads \begin{multline}\label{gl:scatt_int}
\Sigma_{N_\e}\ =\ -\int d\e'\,g(\e)g(\e')\;\times\\
\Big\{W_{\e,\e'}n_\e(1-n_{\e'})-W_{\e',\e}n_{\e'}(1-n_\e)\Big\}\;.
 \end{multline} In this case a relaxation time approximation is not sufficient to describe the experimental data, which are accurate enough to extract the energy dependence of the scattering probabilities $W_{\e,\e'}$.

If inelastic scattering occurs only in the reservoirs, but is negligible in the wire, $\Sigma_{N_\e}$ can be neglected and the resulting $n_\e$ is a two-step function composed of the two Fermi functions of the reservoirs with a weight factor that varies with the position in the wire between 0 and~1. If the inelastic scattering among the electrons is very strong, but negligible between electrons and phonons (relevant for short wires), local thermal equilibrium is re-established and $n_\e$ is a Fermi function with a spatially varying electron temperature $T_\text{el}(r)$. The resulting temperature profile can be determined by solving the thermal diffusion equation with a source term resulting from the local energy dissipation.\cite{steinbach} For very long, but narrow wires the profile of $T_\text{el}(r)$ becomes flat but remains elevated with respect to the phonon temperature, since the electron-phonon coupling becomes very weak under these conditions.

\subsection{Interference of elementary Fermi- and Bose-systems}\label{sec:interference}
So far, we have not taken into account quantum interference between the different modes, or elementary Fermi- and Bose-systems. The effect of interference is simply that two or more modes are not independent anymore, but form {\it new} modes, i.e., the coherent superpositions of the interfering modes. These new elementary sub-systems can be charged with energy, entropy and particles only as one entity, with a transmission coefficient ${\cal T}(\e)$ telling us how constructive or destructive the interference is. The simplest example for such a coherent superposition are the standing waves, which are formed in a quantum well, or a finite quantum wire, by multiple reflection at the confining potential walls. In this case the new systems cannot support a stationary current. Finite currents can be carried only by propagating waves, i.e., scattering 'states', provided that the phases of the interfering waves are adjusted such that the transmission probability is finite. For a completely destructive interference again a standing wave is formed, which suppresses the transport of entropy and particles, and the transport currents impinging on the interferometer are reflected completely. If the source that charges the propagating modes is at a finite temperature, the local current densities of energy, entropy and particles are invariably connected.

The elastic scattering does not produce entropy,  but results in a {\it coherent} branching of the flows of energy, entropy and particle currentsby the scatterer. This statement remains valid for arbitrary complex scattering regions and any number of terminals. The phase coherence can be made visible, if the  partial waves are brought to interference, e.g., by adding a second scatterer or semi-transparent mirror, resulting in an electronic or photonic interferometer of the Fabry-Perot,\cite{zhang} Michelson- or Mach-Zehnder type.\cite{MZI}

For these reasons the possibility of phase-tuning of the transmission coefficients ${\cal T}_k$, and a corresponding phase-dependent thermal conductance\cite{chandra} not very surprising. The close correspondence between the transport of particles and entropy resulting from the present approach renders entropy- or 'heat'-interferometer as natural as a quasiparticle-interferometer, despite the fact that the entropy carried by the elementary Fermi- or Bose-systems results from an {\it incoherent} superposition of states with different particle numbers. The entropy stored within one elementary Fermi- or Bose-system leaves its coherent superposition with others unaffected, as the latter influences only the transmission coefficients, but not the $N_k(T,\mubar)$ and $S_k(T,\mubar)$ responsible for their thermodynamic properties. The phase modulation of the entropy- or 'heat'-transport by optical interferometers must be ubiquitous, but was probably not noted, because the heating power transmitted by single optical modes is limited by the thermal conductance quantum $L_0$ (see Eq.~\ref{gl_term_cond_quantum}), and does not produce large effects at room temperature.

Quantum interference also occurs in diffusive systems. In this case, the modes of the system cannot be chosen as plane waves anymore, but form complex scattering states formed by the interference of waves on multiple scattering centers. The dominating interference contribution to the transport comes from pairs of time reversed diffusion paths, which return to the starting points. The pairs of time reversed paths interfere destructively (at zero external field). The interference increases the probability of return to a given point ($W_\text{return}=1/\sqrt{4\pi Dt}$ for one dimension) by a factor of 2, and thus results in a reduction of the conductivity. This effect, know as weak localization, can be taken into account as an interference contribution to the diffusion constant $D$ with respect to Eq.~\ref{gl:diff_const}. Coherent backscattering of light has also been observed.\cite{montambaux} An external magnetic field tunes the character of the interference between the different diffusion paths in a continuous way between constructive and destructive (Aharonov-Bohm effect).\cite{gantmakher, montambaux} The total transmission probability of a set of interfering modes strongly depends on the wavelength (i.e.,~on $\e$) and via the Aharonov-Bohm effect on the magnetic field $B$ - leading to characteristic conductance fluctuations when $\e_F$ or $B$ are varied.\cite{gantmakher, montambaux}

Another type of interference occurs between pairs of electrons at similar energies. The Coulomb interaction affects the the characteristic energies $\e$ of the elementary Fermi-systems around the Fermi level, and results in a suppression of the DoS near $\e_F$. This effect can be taken into account as a quantum correction to the particle capacity $\nu=\partial n(T,\mu)/\partial \mu$ (Altshuler-Aronov effect).\cite{gantmakher, montambaux}
Similar quantum corrections are also expected in the thermal conductance,\cite{schwab2004} but harder to measure with sufficiently high precision.

In conclusion, the concept of elementary Fermi- and Bose-systems turns out to be extremely flexible. It can be adapted to a wide range of applications in modern physics. Once accepted, it provides a more reliable guide for our intuition than the classical concept of moving particles, as it incorporates the non-classical concepts of quantum interference and indistinguishability from the start.

\subsection{Irreversibility and the Loss of Phase Coherence}\label{irreversibility}
The most ingenious side of the Boltzmann equation is the fact that the aspect of irreversibility, i.e., the generation of entropy is incorporated in the {\it ad-hoc} assumption of the existence of an inelastic relaxation mechanism, and the corresponding characteristic length $\Lambda_\text{in}$. As it was noted very early, such a mechanism is inconsistent with the notion of Hamiltonian dynamics. Any system with a discrete energy spectrum is subjected to the {\it recurrence objection}, i.e., its time evolution must be reversible.
The recurrence objection is removed by assuming the existence of an infinite thermal bath with a continuous spectrum, in which energy and entropy can be dumped without recurrence. Such an approach is successful for systems with a single or a few macroscopic degree of freedom such as quantum bits, coupled to many microscopic degrees of freedom. Irreversibility is then generated by 'tracing out' the bath degrees of freedom.

Phenomenologically, the generation of entropy can be accounted for by damping out the off-diagonal elements of the density matrix, which are responsible for the coherent Hamiltonian dynamics. In the limit of long times, the decoherence becomes complete, implying a density matrix, which is diagonal in the basis of energy eigenstates, and with the probabilities $\{W_i\}$ as eigenvalues. Such a diagonal density matrix corresponds to the state of maximal entropy under the constraints of Eq.~\ref{gl:constraints}, if the $W_i$ are given by Eq.~\ref{gl:Gibbs distribution}.
The joint dynamics of the system and the bath features thermally induced temporal fluctuations of physical quantities of the system. These fluctuations obey the recently much discussed {\it fluctuation theorems}.\cite{hanggi}

A first principles derivation of irreversibility remains a severe problem, as the 'first principles' at hand are all reversible. In the opinion of the author, it is not clear, whether the mathematical operation of 'tracing out the bath degrees of freedom' has a correspondence on the experimental side.
 Moreover, there are situations like the collisions of heavy ions at very high energy, where vast amount of entropy are generated on such short time scales (1~fm/$c\approx 10^{-23}\,$s) that there may exist no sufficiently strongly coupled bath.\cite{high_energy}

\section{Conclusions}
The purpose of this paper is to provide a coherent and self-contained description of the transport of particles and entropy both in the macroscopic and the mesoscopic regime. To implement this program, the concepts of thermodynamics first have to be formulated in a way that avoids unnecessary limitations. In order to connect the general principles of thermodynamics to the quantum physics of matter, the idea of {\it ballistically moving particles} or {\it plane wave propagation}, respectively, has to be stripped from all classical elements. I propose to use the {\it eigenmodes of the matter field} in the language of 2$^\text{nd}$ quantization as the elementary building blocks of such a description. Viewed as thermodynamic systems, called here {\it elementary Fermi- and Bose-systems}, they can be taken as a basis for a unified description of {\it both} global thermodynamic equilibrium {\it and} the ballistic and diffusive quantum transport. In this description 'classical' and quantum transport are the same, while the only demarcation line runs between regimes, where dissipation, i.e., the loss of phase coherence, occurs locally, or remotely in macroscopic reservoirs. The notion of elementary Fermi- and Bose-systems may prove useful not only in solid state physics, including the presently unfolding fields of spintronics, caloritronics, and spin caloritronics, but also in the description of ultracold atomic and molecular gases.\cite{cold_atoms}

\begin{acknowledgments}
This work is dedicated to the late G.~Falk, who introduced me into thermodynamics and its connection to quantum physics. I acknowledge clarifying discussions with M.~Marganska, W. Belzig,  M.~Grifoni, K.~Kang, J.~Siewert, and U.~Sivan. \end{acknowledgments}

\vspace{2mm}

\begin{thebibliography}{99}
\bibitem{note00}
For instance a transfer of a certain amount of heat into a gas during a process does not imply that this heat is now stored within the gas, because a certain amount of energy may be simultaneously extracted as mechanical work. In other words, the gas can convert heat into work. The process quantities heat and work cannot be expressed in terms of the set of independent variables, which characterize the state of the gas. On the other hand, it is correct to say that the entropy transferred during the process is afterwards stored within the gas.
\bibitem{note1a}
The volume is an example of a quantity, which is extensive, but not balanceable, as it has neither a density nor an associated current.
\bibitem{note_maxwell}
The Maxwell relations represent integrability conditions, which ensure that the three EoS can be derived from a thermodynamic potential. They are analogous to those that are obeyed by conservative force fields in mechanics.
\bibitem{footnote3}
Thermodynamic homogeneity has to be sharply distinguished from spatial homogeneity.
The former is defined by Eq.~\ref{gl_hom1} for a set of extensive variables, without reference to spatial coordinates.
\bibitem{footnote5}
A famous exception are the Bogoliubov quasiparticles in a superconductor, where $\qhat$ depends on momentum.
\bibitem{footnote6}
Nevertheless, in solid state physics, $\mubar$ is usually called the chemical potential. This has no big consequences for metals, where the density $n$ is almost fixed (implying that $\mu$ is spatially constant), but matters in semiconductors, where $n$ can vary considerably in space, and the position dependence of $\mu$ cannot be ignored.
\bibitem{footnote7}
This is in some contrast to the common view that thermodynamic rests on the 1.~and 2.~law, which express the conservation of energy, and the impossibility to annihilate entropy. The reason for this is that in many textbooks the quantities work ${\cal W}_\text{rev}$ and heat ${\cal Q}_\text{rev}$ (which are characteristics of reversible processes, rather than states) are considered as fundamental, while changes of energy  $\Delta E={\cal Q_\text{rev}+W_\text{rev}}$ and entropy are $\Delta S={\cal Q}_\text{rev}/T$ are defined by the 1.~and 2.~law. In contrast, in the Gibbsean approach to thermodynamics all the state quantities including $E$ and $S$ are considered as fundamental, while work ${\cal W}_\text{rev}=-\int\limits pdV$ and heat ${\cal Q}_\text{rev}=\int\limits TdS$ are defined in terms of the pairs of state quantities $\{-p,V\}$ and $\{T,S\}$, respectively. See also Ref.~\onlinecite{tisza}.
\bibitem{footnoteZ}
For mobile scatterers, such as the particles in a gas, the relative velocities between the scatterer and the scattered particle may differ from the average velocity of the scattered particles -- this leads to correction factors of the order unity. In many cases, the values of $\sigma_c$ are even less accurately known.
\bibitem{note_j_S_vs j_N}
That for a reversible (i.e., $\Sigma_S=0$ for $T(\vec{r})=const.$) convective entropy flow $\vec{j}_S$ and $\vec{j}_N$ are related by $\partial s(T,n)/\partial n$, and not simply by $\hat{s}=S/N$, is easily seen, if one inserts the continuity equation for the particle density $n$ into the term $\partial s(\vec{r},t)/\partial t = \partial s(T,n)/\partial n\cdot\partial n(\vec{r},t)/\partial t$ within the continuity equation for the entropy density $s$.
\bibitem{footnoteY}
In the literature on correlated electron systems $\nu$ is sometimes called 'thermodynamic density of states'. This terminology is motivated by analogy to ideal degenerate Fermi systems, where  $\partial n(T, \mu)/\partial \mu=g\big(\e_F(n)\big)$. Here $g(\e_F)$ is the density of states at the Fermi level $\e_F$. However this motivation breaks down in dilute Fermi gases like in semiconductors, where $\partial n(T, \mu)/\partial \mu=n/k_BT$, while Eq.~\ref{gl_Einstein} holds in general.
\bibitem{onsager}
In textbooks, the Onsager reciprocity is usually introduced as a consequence of 'microreversibility' referring to the time-reversal symmetry of the scattering events underlying the diffusion process. This reflects the fact that the Hamiltonian dynamics of the scattering can produce correct values for the diffusion constant, without any reference to irreversibility and entropy production (see the discussion in sections~\ref{sec:boltzmann_eq} and \ref{irreversibility}). The present consideration shows that the basic statement of Onsager can be obtained in a very elementary way.
\bibitem{note_Z}
From a mathematical point of view ${\cal Z}$ plays the role of a characteristic function, from which the moments $\langle E^mN^n\rangle$ of the probability distribution can be generated by differentiation with respect to $T$ and $\mubar$. $\ln\,{\cal Z}$ is the the corresponding cumulant function, which produces the cumulants $\langle (E-\langle E\rangle)^m(N-\langle N\rangle)^n\rangle$ of the distribution. This behavior upon differentiation makes $\ln\,{\cal Z}$ suitable as a thermodynamic potential.
\bibitem{note_fluct}
This difference is most obvious for the case of sufficiently dense bosons, where $N_\e$ can beome larger than one.
According to the rules of statistics the variance of these fluctuations is given by the second derivative of $\ln{\cal Z}(T,\mubar)$. The variance of $N_k$ appears experimentally as thermal {\it noise}. This is an important conceptual change with respect to the Boltzmann view, in which particles are distributed with probabilities $W_k$ over single particle states $|k\rangle$. Probabilities are not random numbers -- they do not fluctuate, even if in Fermi-systems the numerical values of the $N_k$ and the probabilities $W_{k1}$ for fermion states $|k\rangle$ with one particle are the same! The Boltzmann approach works in the dilute limit only, where $N_\e\ll1$ can indeed be interpreted as probabilites, and the difference between fermions and bosons is not important.
\bibitem{footnote_k_e}
The transition from discrete $\{k\}$- to continuous $\{\e\}$-values as labels for the elementary Bose- or Fermi systems masks the implicit dependence of the allowed $k$ and $\e$ values on the volume. For simplicity we will suppress it in the following, but the reader should be aware that this dependence on $V$ determines the Gr\"{u}neisen parameter, e.g., the thermal expansion coefficient of the system (see Ref.~\onlinecite{ashcroft}).
\bibitem{footnote_a}
In the case of ultracold gases the potential well is often a parabolic trap, resulting in a harmonic oscillator potential with its corresponding equidistant energy eigenvalues $\e_i=\hbar\omega(i+1/2)$.  As the oscillator wave functions with different $\e_i$ differ in their spatial extension, the elementary subsystems do not cover the same volume in space, but the concepts of the elementary Fermi- and Bose-systems remain unchanged.
\bibitem{note_reservoirs}
In reality, this basic assumption of the Landauer-B\"{u}ttiker approach can be at best approximatively true. Its validity depends on whether the fraction of non-equilibrium quasi-particles backscattered from the reservoirs into the quantum wire is is small enough, to avoid a significant deviation of the left-mover's $N_{\e;\,L,R}$ from their reservoir values.
In other words, the inelastic relaxation in the reservoirs has to be sufficiently efficient. This condition that can be violated in practice, in particular if the elastic mean free path in the reservoirs is short, or if the $\mubar$ and/or $T$-differences are large.
\bibitem{note_nonlinear}
This discrepancy originates from the fact that local values of $T$ and $\mubar$ cannot be defined in the nonlinear regime, and thus Eq.~\ref{gl_GFF_J} is invalid.
\bibitem{note_rel_time_approx}
This refinement is kwon as the relaxation time approximation.\cite{ashcroft} It is still a rather crude approximation, as it assumes that the scattering time is independent of the particle numbers in the other elementary Fermi- or Bose-systems.
\bibitem{footnote_L0}
The most powerful methods for the calculation of the transmission coefficient for complex scatterers are provided by the  Green's-function approach to quantum field theory, which allows to take into account all the quantum interference mentioned in Sec.~\ref{sec:interference} and below. From the conceptional point of view these methods do not go very much beyond of our elementary approach.
\bibitem{footnote_LB}
This holds in the linear regime, where an effect of the bias voltage on the set $\{{\cal T}(\e)\}$ can be neglected.
\bibitem{note10}
This randomization is not complete (otherwise there would be no net electric current), but the remaining anisotropy is so small that local equilibrium can be assumed as long as the gradient of the electrochemical potential $|\nabla \mubar|=\qhat U/L$ remains small compared to $k_BT/\Lambda_\text{el}$, where $\Lambda_\text{el}$ is the elastic mean free path.
\bibitem{note11}
From the point of view of classical mechanics, it is temp\-ting to say that particles emanating from the same reservoir maintain their thermal and chemical equilibrium with that reservoir during the diffusion process --  but this is impossible as the indistinguishability of electrons prevents to record the 'history', i.e., the diffusion path of one particular particle.
\end{thebibliography}

\end{document}